%% file: for.tex
\documentclass[usenatbib]{mn2e}

\usepackage{arydshln}

\input{commands}

\begin{document}

\title[Combining surveys II: \tfor]{\tserieslong\ II: \tfor}
\author[Martin Eriksen, Enrique Gazta\~naga]
{\parbox{\textwidth}{Martin Eriksen$^{1,2}$, Enrique Gazta\~naga$^1$}
\vspace{0.4cm}\\
$^1$Institut de Ci\`encies de l'Espai (IEEC-CSIC),  E-08193 Bellaterra (Barcelona), Spain \\
$^2$Leiden Observatory, Leiden University, PO Box 9513, NL-2300 RA Leiden, Netherlands \\
}

\maketitle
\begin{abstract}
\input{abstr}
\end{abstract}

\section{Introduction}
\input{intro}

\section{Forecast assumptions}
\label{assumpt}
\input{assumpt}

\section{Results}
\input{results}


\section{Conclusion}
\input{concl}

\section*{Acknowledgments}
\input{acknow}

\appendix
\section{Impact of non-linear scales}
\label{app_nonlin}
\input{nonlin}

\section{Contours}
\label{app_contours}
\input{contours}

\bibliographystyle{mn2e}
\bibliography{../exbib}{}
\end{document}

%% file: commands.tex
\usepackage{amsmath}
\usepackage{graphicx}
\usepackage{color}
\usepackage{aas_macros}
\usepackage{float}

\newcommand{\width}{0.5\textwidth}
\newcommand{\tserieslong}{Combining spectroscopic and photometric surveys
using angular cross-correlations}

\newcommand{\tfor}{Parameter constraints from different physical effects}

\newcommand{\gkone}{FxB-$\left<\delta_F \gamma_F\right>$}
\newcommand{\gktwo}{FxB-$\left<\delta_B \gamma_F\right>$}
\newcommand{\gkthree}{FxB-$\left<\delta \: \gamma \right>$}
\newcommand{\gkthreens}{FxB-$\left<\delta \gamma \right>$}

\newcommand{\gkfb}{FxB-$\left<\text{FB}\right>$}

\newcommand{\nbright}{71}

\newcommand{\be}{\begin{equation}}
\newcommand{\ee}{\end{equation}}
\newcommand{\xbf}{\begin{figure}}
\newcommand{\xef}{\end{figure}}
\newcommand{\xbfd}{\begin{figure*}}
\newcommand{\xefd}{\end{figure*}}
\newcommand{\xbt}{\begin{table}}
\newcommand{\xet}{\end{table}}

\newcommand{\citedir}{\citet}
\newcommand{\citeind}{\citep}

\newcommand{\tmpcitetwo}[1]{}
{}
\newcommand{\tmprm}[1]{}
\newcommand{\xfigure}[1]{
\begin{center}
\includegraphics[width=\width]{links/#1}
\end{center}
}

\newcommand{\micecosmo}{$\Omega_m = 0.25$, $\Omega_b = 0.044$, $\Omega_{DE} =
0.75$, $h = 0.7$, $w_0 = -1$, $w_a = 0$, $n_s = 0.95$ and $\sigma_8 = 0.8$ }

\newcommand{\dzmax}{\Delta Z_ \text{Max}}
\newcommand{\dzbin}{\Delta z}

\newcommand{\deltal}{\Delta l}

\newcommand{\kmax}{k_{\text{max}}}
\newcommand{\fsky}{f_\text{Sky}}
\newcommand{\sqarcmin}{\text{arcmin}^2}

\newcommand{\hinv}{\text{h}^{-1}}
\newcommand{\mpc}{\text{Mpc}}

\newcommand{\fom}{\text{FoM}_{w}}
\newcommand{\fomg}{\text{FoM}_\gamma}
\newcommand{\fomc}{\text{FoM}_{w\gamma}}
\newcommand{\fomdetf}{\text{FoM}_{\text{DETF}}}

\newcommand{\citeoverlap}{\citeind{cai2011,gazta,cai2012,kirk,mcdonald,deputter}}


%% file: abstr.tex
Future spectroscopic and photometric surveys will measure accurate positions
and shapes of an increasing number of galaxies. In the previous paper of this
series we studied the effects of Redshift Space Distortions (RSD), baryon
acoustic oscillations (BAO) and Weak gravitational Lensing (WL) using angular
cross-correlation. Here, we provide a new forecast that explores the
contribution of including different observables, physical effects (galaxy bias,
WL, RSD, BAO) and approximations (non-linearities, Limber approximation,
covariance between probes). The radial information is included by using the
cross-correlation of separate narrow redshift bins. For the auto correlation
the separation of galaxy pairs is mostly transverse, while the
cross-correlations also includes a radial component. We study how this
information adds to our figure of merit (FoM), which includes the dark energy
equation of state $w(z)$ and the growth history, parameterized by $\gamma$. We
show that the Limber approximation and galaxy bias are the most critical
ingredients to the modelling of correlations. Adding WL increases our FoM by
4.8, RSD by 2.1 and BAO by 1.3. We also explore how overlapping surveys perform
under the different assumption and for different figures of merit. Our
qualitative conclusions depend on the survey choices and scales included, but
we find some clear tendencies that highlight the importance of combining
different probes and can be used to guide and optimise survey strategies.

%% file: intro.tex
The expansion of the universe provide a challenge for cosmology and fundamental
physics. Understanding the recent accelerated expansion of the universe is
connected to dark matter and dark energy, either by determining their
properties or by providing an alternative theory. There is no scarcity of
models, but no model beyond $\Lambda$CDM has emerged as a natural candidate to
explain the cosmic acceleration.

Galaxy surveys are designed to probe cosmology in different manners. Weak
lensing of foreground matter affects the galaxy shapes. Observing the weak
lensing through galaxy shapes (shear) \citeind{shearcosmos,shearcosmos2}
requires deep imaging surveys like DES and the upcoming Euclid and LSST.
Further, overdensities of dark matter attracts nearby galaxies, which creates
an additional peculiar velocity. The radial component of the extra velocity
results in a shift in redshift, which is the effect of redshift space
distortions (RSD). Optimal measurement of RSD require accurate redshift and
the method is most suitable for spectroscopic surveys.

The lensing efficiency has a broad kernel and the shear-shear lensing signal
can be analyzed in 5-10 broad redshift bins. The RSD signal is traditionally
analyzed using  power spectrum analysis in 3D comoving space, which includes a
cosmology dependent conversion transformation of angles and redshift to the 3D
comoving distances. As shown in paper-I (and references therein), angular
correlations can also be used to measure RSD. A previous study \citedir{asorey}
found that angular correlations in narrow redshift bins can recover most of the
information in the 3D power spectrum. In these papers we use the angular
correlation for both the photometric (WL) and spectroscopic (RSD, WL) survey.

Using a single set of observables for WL and RSD has several advantages. For
overlapping surveys the galaxies trace the same matter fluctuations, which
introduce a covariance between the surveys. Particular care is needed to not
double counting information when jointly observing shear-shear in the lensing
survey, the 3D power spectrum for the spectroscopic survey and 2D correlations
for the counts-shear cross-correlations between the two. For example counting
the modes is insufficient if including photo-z effects in the 3D sample since
the photo-z affects the radial and transverse modes differently. 

Several groups have explored combining weak lensing and spectroscopic surveys
\citeoverlap. In overlapping surveys (same-sky) one can cross-correlate the
observables, e.g. galaxy counts from the two surveys or galaxy counts from the
spectroscopic sample with background shear from the lensing survey.
Overlapping samples further reduce the sample variance \citeind{mcdonseljak}.
While non-overlapping surveys benefits from larger area, several authors find
stronger parameter constraints when combining weak lensing and spectroscopic
surveys over the same area. In this paper we study the importance of different
physical effects for overlapping and non-overlapping surveys, while the paper
\citedir{sameskyX} elaborate on the benefit of overlapping surveys.

Galaxies are theoretically \citeind{hod1} and observationally expected to form
in overdense regions. Unlike shear which is affected by all foreground matter,
the galaxy counts relates to the underlying matter distribution at a given
redshift. The negative aspect of probing cosmology with the galaxy counts 
is requiring to understand and marginalise over uncertainties in the relation
between matter and galaxy overdensities, the galaxy bias \citeind{gazta}. One
can either approach the galaxy bias by only using the BAO peak
\citeind{seoeisenstein,seoeisenstein2} or parameterize the bias
\citeind{fullcls}. In this paper we use the full galaxy correlation and measure
the bias parameters by combining LSS and lensing in a multiple tracer
analysis.

Magnification change the overdensities of number counts through two weak
lensing effects. In a magnitude limited sample, lensed galaxies appear brighter
and enters into the sample when magnified over the magnitude cut. The
magnification also magnifies the area which reduces the number density. In the
SDSS sample magnification has been observed by correlating foreground galaxies
with background quasars \citeind{scrantonmag,menard2}. While the shear-shear
signal has less noise, magnification provides an additional signal which is
already present in the galaxy catalogs. This paper, in a similar way to
\citedir{gazta,duncanmag}, will study the benefits of magnification when
combining the analysis of spectroscopic and photometric surveys in angular
correlations.

Photometric surveys conventionally use broad band filters. Two upcoming
surveys, PAU\footnote{www.pausurvey.org} and J-PAS\footnote{j-pas.org} plan to
measure photometry for galaxies in 40 to 50 narrow (100-130A) bands. For PAU
the resulting photo-z precision is $\sigma_{68} \approx 0.0035(1+z)$ for
$i_{AB} \leq 22.5$ \citeind{polpz}. In addition the PAUcam broad bands (ugrizY)
has an anticipated photo-z accuracy of $\sigma_{68} = 0.05(1+z)$ for $22.5 <
i_{AB} < 24.1$. The PAU survey at the William Herschel Telescope (WHT) can
cover about  200 sq.deg. to $i_{AB} < 22.5$ in narrow bands and  $i_{AB} < 24$
with broad bands in 100 nights. These defines two magnitude limited populations
of Bright (B) galaxies ($i_{AB} < 22.5$) and Faint (F) galaxies ($22.5 < i_{AB}
< 24$) similar to two overlapping spectroscopic and photometric surveys.

This paper (paper-II) is part of a three paper series. The first paper
(paper-I, \citedir{paperI}) dealt with modelling of of the correlation function,
focusing in particular on the effect of RSD, BAO and the Limber approximation
in narrow bins. Here (paper-II) we forecast the relative impact of WL, RSD and
BAO in upcoming cosmological surveys. A third paper (paper-III,
\citedir{paperIII}) study the impact of galaxy bias, while a separate paper
\citeind{sameskyX} focus on the benefit of overlapping surveys.

This paper is organised in the following manner. The second section present the
assumptions, which includes Fisher matrix formalism, forecast assumptions and
nomenclature. In the third section we compare the relative contribution of the
different effect (including WL, RSD, BAO, magnification) and the Limber
approximation. Last section is the conclusion.

%% file: assumpt.tex
This section first present the assumptions fiducial cosmology, galaxy bias
parameterization, galaxy samples, surveys definitions, cuts in non-linear
scales and the Fisher forecast. Paper-I included the theoretical expressions
for the $C_l$ cross-correlations and they are therefore not repeated here.  In
this section we also define the FoMs and the nomenclature (e.g. FxB, F+B) used
throughout the paper.

\subsection{Fiducial cosmological model.}
\newcommand{\lcdm}{$\Lambda$CDM}

The cosmological model assumed is $w$CDM\footnote{
For the fiducial cosmology we use the values \micecosmo , which correspond to
the cosmological model in the MICE (http://www.ice.cat/mice) simulation.},
which is General Relativity with Cold Dark Matter and a dark energy model with
an equation of state $w \equiv \frac{p_{DE}}{\rho_{DE}}$. The main observables
are $C_l$ cross-correlations of fluctuations $\delta(z,k)$, which depends on
the initial power spectrum, distances and the growth of fluctuations. For a
Friedmann-Lema\^{i}tre-Robertson-Walker (FLRW) metric \citeind{dodelson} , the
Hubble distance is

\begin{align}
H^2(z) &= H_0^2 [\Omega_m a^{-3} + \Omega_k a^{-2} + \rho_{DE}(z)] \\
\rho_{DE} &= \Omega_{DE} a^{-3(1+w_0 + w_a)} 
  \exp{\left(-3w_a z/(1+z)\right)}
\label{hubble_law}
\end{align}

\noindent
where the last equation expresses the dark energy density using the 
parameterization

\be
w(z) = w_0 + w_a(1-a)
\ee

\noindent
from \citedir{linder, linderparam} for the dark energy equation of state
(EoS). Overdensities of matter grow because of gravitational attraction  and at
large (linear) scales the equation determining the growth has the solutions
\citeind{heath1977, peebles}

\be
\delta(z) = D(z) \delta(0)
\label{grow_def}
\ee

\noindent
where here $D(z)$ is defined through

\be
f \equiv \frac{d \ln(D)}{d \ln(a)} = \frac{\dot{\delta}}{\delta} \equiv \Omega_m^\gamma(a)
\label{gamma_def}
\ee

\noindent
Normalising the growth to $D(z=0)=1$ we have

\begin{equation}
D(z) = \exp\left[ -\int_a^1 d\ln{a} f(a) \right].
\label{growth}
\end{equation}

\noindent
In these papers (and previously in \citedir{gazta}) the growth is parameterized
through the parameter $\gamma$  in Eq.\ref{gamma_def}, which is $\gamma \approx 3/11
\approx 0.55$ in General Relativity with a cosmological constant. For example
the DGP model \citeind{dgp} propose to explain the cosmological acceleration
through embedding the ordinary 3+1 dimensional Minkowski space in a 4+1
dimensional Minkowski space. Alternatively modified gravity, which we have left
of future work, can be parameterized by the Bardeen potentials
\citeind{bardeen}. Adjusting free parameters in modified gravity can
potentially fit the right expansion history (Eq. \ref{hubble_law}), but it is
more difficult to simultaneously fit the expansion and growth history (Eq.
\ref{growth}). Constraining both the growth and expansion history is therefore
important to discriminate between different modified gravity models.

\subsection{Non-linear scales}
\label{subsec_nonlin}
On lager scales fluctuations are linear. In contrast, for high density regions
the structures collapses in a non-linear manner. As a result, predicting the
non-linear power spectrum require either simulations \citeind{gadget},
perturbation theory \citeind{pertcrocce} or fitting functions to simulations
\citeind{coyote1,coyote2,coyote3}. Here the forecast use the Eisenstein-Hu
\citeind{eishu} linear power spectrum. In appendix \ref{app_nonlin} we
test the effect of  including the non-linear contribution.

Even when including non-linear power spectrum contributions, one need to limit
the maximum $k_{max}$ (or minimum $r_{min}$) scale.  The Halofit II model is
calibrated to 5\% percent accuracy for $k \leq 1 h \mpc^{-1}$ at $0 \leq z \leq
10$. Further, we also want to limit observations to scales where the bias (see
subsection \ref{subsec:bias}) is scale independent. Here and in other papers of
these series the maximum scale is defined through

\be
\sigma(R_{\text{min}}, z) = 1
\ee

\noindent
where $\sigma(R, z)$ is the fluctuation amplitude smoothed with a Gaussian
kernel on a scale $R$. From $k = l/ \chi(z)$, we have

\be
\kmax(z) = \frac{R(0)_{\text{min}}}{R(z)_{\text{min}}} \kmax(0)
\ee

\noindent
where $\kmax(0) = 0.1 \hinv \text{Mpc}$ is an overall normalisation. In the
MICE cosmology and Eisenstein-Hu power spectrum, then

\be
k_{\text{max}}(z) = \exp(-2.29 + 0.88 z)
\label{kmax_crit_fit}
\ee

\noindent
is a good fit for the $\kmax$ limit. The conversion to an $l_{\text{max}}$ for
which correlations to include is done with

\be
k_{\text{max}} = \frac{l_{\text{max}} + 0.5}{r(z_i)}
\label{kmax_crit_fit2}
\ee

\noindent
which uses the scale contributing to LSS and counts-shear correlations in the
Limber equation (see paper-I). For cross-correlations we use the minimum
$k_{max}$ from the two redshift bins. The forecast is restricted to $10 \leq l
\leq 300$ and in addition apply the cut above, including for the the
shear-shear correlations. To save time, the forecast use $\Delta l = 10$. We
have tested that the discrete l-values have minimal impact on the forecast.

\subsection{Galaxy bias}
\label{subsec:bias}
Galaxy overdensities $\delta$ are in a local bias model \citeind{frybiasmodel} related to
matter overdensities $\delta_m$ through

\begin{equation}
\delta(k,z) = b(z,k) \delta_m(k,z)
\end{equation}

\noindent
where the bias $b(k, z)$ can in general depend on scale and redshift. Each
subset of galaxies, of galaxy population, can have different bias since galaxy
types (e.g. elliptical and spirals) cluster and evolve differently. When
defining populations by magnitude cuts, as we do, the bias also differ because
each population contains another mixture of galaxies.

The two galaxy populations (see subsection \ref{subsec:fidsurveys}) use a
different bias and bias nuisance parameters. We use one bias parameter per
redshift bin and galaxy population, no scale dependence and no additional bias
priors. In addition, the bias can include a stochastic component. A common used
measure of non-linearity and the stochasticity is

\be
r \equiv \sqrt{\frac{C_{\delta m}}{C_{\delta \delta}C_{\delta m}}}
\ee

\noindent
where $C_{\delta \delta} \equiv \left<\delta\delta\right>$, $C_{\delta
\delta_m}  \equiv \left<\delta m\right>$ and $C_{mm}  \equiv \left<\delta_m
\delta_m \right>$ respectively are the counts-counts, counts-matter and
matter-matter correlations. For a deterministic and linear bias, then $r = 1$.
In  \citedir{gazta} we showed by theoretical models and also simulations that
the stochasticity can be treated as a re-normalisation of the bias. Thus we fix
the stochasticity to $r=1$ and explore the impact of $r$ in paper-III.

\subsection{Fisher matrix forecast}
The Fisher matrix is a simple and fast method to estimate parameter
uncertainties. Deriving the Fisher matrix follow from a Gaussian approximation
of the Likelihood expanded around the fiducial value. Sampling the Likelihood
with MCMC methods would be more precise, but greatly increase the computation
time. Since the Fisher matrix is widely used in the literature, including the
results we compare with, we also use the Fisher matrix formalism.

For the correlations $C_{ij}$ and corresponding covariance Cov, the Fisher
matrix is

\begin{equation}
F_{\mu \nu} = \sum_{ij,kl} \frac{\partial C_{ij}}{\partial \mu} 
  {\text{Cov}}^{-1} \frac{\partial C_{kl}}{\partial \nu}
\label{fisher_mat_formula}
\end{equation}

\noindent
where $\mu$ and $\nu$ are parameters and the two sums are over different
correlations.  If the observable does not enter the forecast, it is not
included neither in the sums nor the covariance. One example of dropping
observables is the removal of non-linear scales, as explained in the last
subsection. The Cramer-Rao bound states that

\begin{equation}
\sigma^2_{\mu} \leq F_{\mu \mu}^{-1}
\end{equation}

\noindent
where $F^{-1}$ denote the Fisher matrix inverse and $\sigma_{\mu}^2$ the
expected parameter variance for the parameter $\mu$. The covariance matrix of
2D-correlations when assuming Gaussian fluctuations \citeind{dodelson} is

\begin{equation}
Cov(C_{AB}, C_{DE}) = N^{-1}(l) (C_{AD} C_{BE} + C_{AE} C_{BD})
\label{covariance}
\end{equation}

\noindent
where number of modes is $N(l) = 2 \fsky (2l + 1) / \deltal$, $\fsky$ is the
survey fractional sky coverage and $\deltal$ is the band width (bin width in
l).

Adding constrains from uncorrelated observables
require summing up the Fisher matrices. For example the constraints of LSS/WL
and CMB is

\be
F_{\text{Combined}} = F_{\text{LSS/WL}} + F_{\text{CMB}}
\label{fadd}
\ee

\noindent
when assuming the CMB is sufficiently uncorrelated with the LSS/WL experiment.
One can prove Eq.\ref{fadd} using the covariance for two uncorrelated set of
parameters is block diagonal and Eq.\ref{fisher_mat_formula} can be split in
two parts. For the forecasts all results (unless explicitly stated) add
Planck priors\footnote{We use the Planck Fisher matrix file "planckfish" from
http://www.physics.ucdavis.edu/DETFast/.}.

\subsection{Figures of Merit (FoMs)}
\label{subsec:assum_fom}

Figure of Merits are a simplified representation of the parameter constraints.
A Fisher matrix of $n$ parameters includes $n (n+1)/2$ independent entries.
Instead of including all the information on the errors and the covariance
between parameters present in a covariance matrix, the figure of merit is only
a single number. Comparing probes, effects and configurations are greatly
simplified when using a single number. The FoMs let us study the gradual change
with a parameter given on the x-axis, while adding different lines
corresponding to various configurations. Also, the FoM is useful for comparing
information along two dimensions in a table. While a single number does not
fully capture the utility of a galaxy survey, but is a good measure to discuss
trends.

The parameters included in the Fisher matrix forecast are
\begin{equation*}
w_0, w_a, h, n_s, \Omega_m, \Omega_b, \Omega_{DE}, \sigma_8, \gamma, \text{Galaxy Bias}
\end{equation*}

\noindent
where the 9 first parameters equals the one included in the dark energy task
force (DETF) figure of merit \citeind{detf}. The galaxy bias (see subsection
\ref{subsec:bias}) is parameterized with one parameter in each redshift bin for
both galaxy populations. Fiducially, this study ignore the bias stochasticity,
shear intrinsic alignments \citeind{hirata,intrshear}, uncertainties in photo-z
distributions \citeind{newman,newman2} and other shear systematics
\citeind{bernshear}.

The DETF figure of merit is inversely proportional to the
$(w_0, w_a)$ 1-$\sigma$ contour area. Analogous \citedir{gazta} defined an
extended FoM,

\begin{equation}
\text{FoM}_S \equiv \frac{1}{\sqrt{\det\left[F_S^{-1}\right]}}
\end{equation}

\noindent
where S is a parameter sub-space. Parameters not in $S$ are marginalised over.
Since this concept is quite natural, other papers (e.g. \citedir{asorey,
kirk,asorey}), define similar FoMs. Identical to \citedir{gazta}, we define
three FoMs (in addition to  DETF FoM)

\begin{itemize}
\item $\fomdetf$. $S=(w_0, w_a)$. Dark Energy Task Force (DETF) figure of merit. 
  Inversely proportional to the error ellipse of ($w_0$, $w_a$).

\item $\fom$. $S=(w_0, w_a)$. Equivalent to $\fomdetf$, but instead of 
  $\gamma = 0.55$ from GR, the $\gamma$ is considered a free parameter
and is marginalised over.

\item $\fomg$ $S=(\gamma)$. Inverse error on the growth parameter $\gamma$, when 
  marginalising over the other cosmological parameters and the galaxy bias. Therefore e.g.
  $\fomg = 10,100$ respectively corresponds to 10\%,1\% expected error on $\gamma$.

\item $\fomc$. $S=(w_0, w_a, \gamma)$. Combined figure of merit for $w_0$, $w_a$ and $\gamma$. 
  The 3D determinant also includes the correlation between the dark energy
  $(w_0, w_a)$ and growth ($\gamma$) constraints.
\end{itemize}

\noindent
Note that different authors introduce numerical prefactors of 1/4 (or 1/$4\pi$) 
\citeind{bridleking,joachbridle} in the FoM. In these papers results are often
presented in $\fomc$, while the other FoMs are used to disentangle gains in
measuring expansion and growth history. One should be aware that the FoMs scale
with area in the following way

\begin{align}
\nonumber
\fom &\propto A \\
\nonumber
\fomdetf &\propto A \\
\nonumber
\fomg &\propto A^{1/2} \\
\fomc &\propto A^{3/2} 
\label{areagain}
\end{align}

\noindent
when not including priors and $A$ is the survey area. Including prior reduce
the slope for small areas if the prior dominate. From numerical tests (not
shown), the scaling above works well for the fiducial 14.000 sq. deg. survey.

\subsection{Fiducial galaxy surveys}
\label{subsec:fidsurveys}
The two defined populations corresponds to a spectroscopic (Bright) and
photometric (Faint) survey. Both populations are magnitude limited, although a
spectroscopic survey often would select specific targets to optimise the science
return. The fiducial area is 14.000 sq.deg., which is around the expected sky
coverage of stage-IV surveys.

\xbf
\xfigure{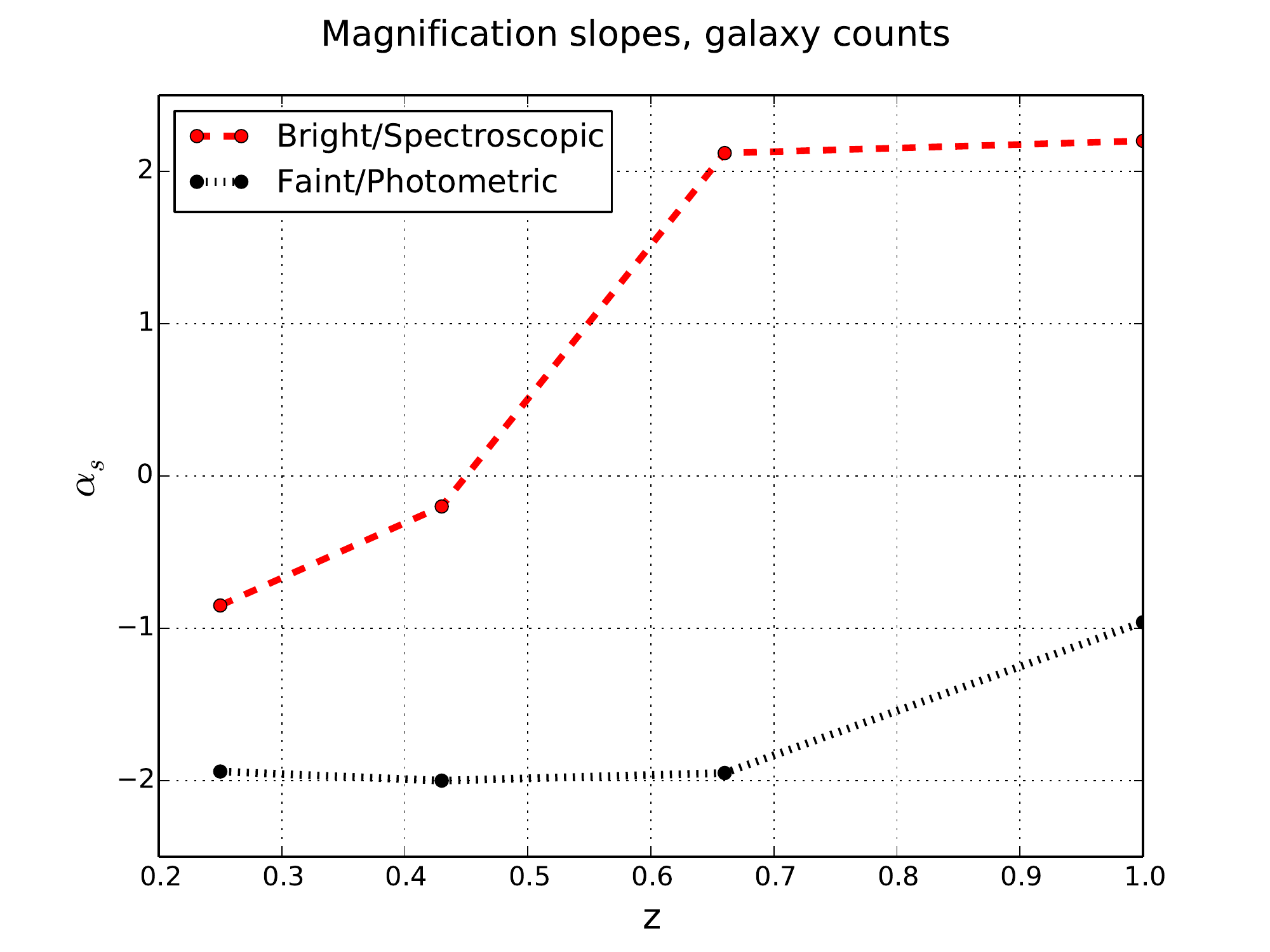}
\caption{The number counts magnification slopes ($\alpha_s$) for a Bright ($18
< i_{AB} <= 22.5$) and Faint ($22.5 < i_{AB} < 24.1$) from the COSMOS sample.
Values are extrapolated outside of this redshift range.}
\label{mag_slopes}
\xef

Properties of the two populations, are defined in the next two subsections,
with Table \ref{tbl_assumpt} summarising the central values. The shape of the
galaxy distributions (not density) and also galaxy bias corresponds exactly to
the values in \citedir{gazta}. There the galaxy distributions was constructed
by fitting a Smail type n(z) \citeind{smail}

\newcommand{\txtnz}[1]{\frac{dN_{\text{#1}}}{d\Omega dz}}
\newcommand{\nz}[4]{A_{\text{#1}} {\left(\frac{z}{#2}\right)}^{#3} \exp{\left(-{\left(\frac{z}{#2}\right)}^{#4}\right)}}

\be
\txtnz{F} \propto \nz{F}{z_0}{\alpha}{\beta}
\ee
to the public COSMOS photo-z sample. In addition the magnification (see
paper-I) adds the term $\delta^{WL} = \alpha_s \delta_{\kappa}$, where $\kappa$
is the convergence, to the galaxy counts overdensities. Fig. \ref{mag_slopes}
specify the fiducial magnification slopes ($\alpha_s$).

\newcommand{\tblskipa}{\noalign{\vskip 1.0mm}}
\newcommand{\tblmidlinea}{\tblskipa\hdashline\tblskipa}

\begin{table}
\begin{center}
\vskip+0.5ex
\begin{tabular}{lrr}
\hline
Parameter & Photometric (F) & Spectroscopic (B)\\
\hline
Area [sq.deg.] & 14,000 & 14,000 \\
Magnitude limit & $i_{AB} < 24.1$& $i_{AB} < 22.5$ \\
Redshift range & $0.1 < z < 1.5$ & $0.1 < z < 1.25$ \\
Redshift uncertainty & 0.05(1+z) & 0.001 (1+z) \\
z Bin width & 0.07 (1+z)& 0.01(1+z) \\
Number of bins & 12 & \nbright \\
\tblmidlinea
Bias: b(z) & 1.2 + 0.4(z - 0.5)  & 2 + 2(z - 0.5) \\
Shape noise & 0.2 & No shapes \\
\tblmidlinea
density [gal/arcmin$^2$] &  6.5 & 0.4 \\
nz - $z_0$  & 0.702 & 0.467  \\
nz - $\alpha$ & 1.274 & 1.913 \\
nz - $\beta$ & 2.628 & 1.083 \\
\hline
\end{tabular}
\end{center}
\caption{Parameters describing the two surveys/populations. The first section
give the area, magnitude limit, redshift range used in the forecast, redshift
uncertainty modelled as a Gaussian, the redshift bin with and the resulting
number of bins. In the second section is the galaxy bias ($\delta = b
\delta_m$) and average galaxy shape uncertainty. The third section give the
galaxy density and parameters for the n(z) shape.}
\label{tbl_assumpt}
\end{table}

\subsubsection{Bright/Spectroscopic population}
The Bright population is defined by the flux limit $i_{AB} < 22.5$, has a
Gaussian spectroscopic redshift uncertainty of $\sigma_{68} = 0.001(1+z)$ and
the galaxy density 

\be
\txtnz{B} = \nz{B}{0.702}{1.083}{2.628}
\ee

\noindent
over $0.1 < z < 1.2$. Here $A_B$ is a normalisation amplitude and the fiducial
density is 0.4 gal/$\sqarcmin$, which is dense for a spectroscopic survey.  The
redshift evolution of the galaxy count bias is defined by

\be
b_B(z) = 2 + 2 (z - 0.5)
\ee
where $b(0) = 1$. To recover the radial information in the Bright/spectroscopic
sample, we use $\Delta z = 0.01 (1+z)$ narrow redshift bins.

\subsubsection{Faint/Photometric population}
Weak gravitational lensing require a dense and deep sample with imaging to
measure galaxy shapes. The Faint population resemble a wide field lensing
survey, magnitude limited to $i_{AB} < 24.1$ and $\sigma_{68} = 0.05 (1+z)$
Gaussian photo-z accuracy. This deeper magnitude selection give the galaxy
distribution:

\begin{equation}
\txtnz{F} = \nz{F}{0.467}{1.913}{1.274}
\end{equation}

\noindent
over $0.1 < z < 1.4$. The complete Faint density is 17.5 gal/sq.arcmin and in
addition we only use 50\% of the galaxies, which can either come from photo-z
quality or shear measurements cuts. Similar to the spectroscopic sample, the
bias model is linear with

\begin{equation}
b_F(z) = 1.2 + 0.4 (z - 0.5)
\end{equation}

\noindent
which also has $b(0) = 1$. The Faint sample use $\Delta z = 0.07 (1+z)$ thick
redshift bins and contribute strongest to the WL constraints. Decreasing the
bin width would not improve the radial resolution or the RSD signal, as the
photo-z error introduce an effective binning in redshift \citeind{gazta}.

\subsection{Observables}
\label{subsec:assum_observ}

\input{observ_tbl}

\newcommand{\tblspacec}{\noalign{\vskip 2.0mm}}
\newcommand{\colr}{0.8}
\begin{table}
\begin{center}
\begin{tabular}{ll}
\hline
Notation & Description \\
\hline 
$\left<\delta_F\delta_B\right>$ &
\begin{minipage}[t]{\colr\columnwidth}
Counts-Counts cross-correlations of the two galaxy populations. Important
for sample variance cancellation. \end{minipage}\\
\tblspacec  
$\left<\delta_B\gamma_F\right>$ &
\begin{minipage}[t]{\colr\columnwidth}
Counts-Shear cross-correlations of foreground spectroscopic galaxy counts
and shear.\end{minipage} \\
\tblspacec  
$\left<\gamma_B \gamma_B\right>$ &
\begin{minipage}[t]{\colr\columnwidth}
Shear-Shear for the Bright galaxies. In overlapping surveys the Bright galaxies
is a subset of the photometric survey. This term is of minor importance since
the Bright sample is shallower and less dense. \end{minipage} \\
\hline 
\end{tabular}
\end{center}
\caption{Notation for different cross-correlations. First column give the
cross-correlation and the second column is a short description.}
\label{notation_two}
\end{table}

Table \ref{notation_one} define the notation and also give the observables
included for a list of different cases. In this series of papers, a main topic
is on the combined constraints from photometric (F) and spectroscopic (B)
surveys, either alone (F or B) or for overlapping (FxB) or non-overlapping
(F+B) areas.  The "All" notation means both shear and galaxy counts, while
"Counts" only includes counts. Part of the benefit of overlapping surveys come
from additional cross-correlations. To quantify their impact, the second part
of the table therefore present the notation for removing selected
cross-correlations. Table \ref{notation_two} contains a list of cross-correlations
which enters both directly in the text and Table \ref{notation_one}.

%% file: observ_tbl.tex
\newcommand{\tblpadb}{\noalign{\vskip 1.0mm}}
\newcommand{\tblskipb}{\noalign{\vskip 1.3mm}}
\newcommand{\tblspaceb}{\noalign{\vskip 1.0mm}}
\newcommand{\tblspacebb}{\noalign{\vskip 2.2mm}}
\newcommand{\tblmidlineb}{\tblskipb\hdashline\tblskipb}

\newcommand{\mywidth}{0.75\columnwidth}
\begin{table*}
\begin{center}
\begin{tabular}{l l l}
\hline
Notation & Observables & Description \\
\hline
\tblpadb
F:All 
 & 
\begin{minipage}[t]{\mywidth}
$\left<\delta_F \delta_F\right>+\left<\delta_F \gamma_F\right>+\left<\gamma_F \gamma_F\right>$
\end{minipage} 
 & 
Faint population
\\
\tblspaceb
B:All 
 & 
\begin{minipage}[t]{\mywidth} $\left<\delta_B
\delta_B\right>+\left<\delta_B \gamma_B\right>+\left<\gamma_B \gamma_B\right>$
\end{minipage} 
&
Bright population
\\
\tblspaceb
FxB:All
 &
\begin{minipage}[t]{\mywidth} F+B:All + $\left<\delta_F
\delta_B\right>+\left<\delta_B \gamma_F\right>$ \end{minipage} 
&
Overlapping Faint and Bright.
\\
\tblspaceb
F+B:All
 &
\begin{minipage}[t]{\mywidth}
F:All + B:All -  $\left<\gamma_B \gamma_B \right>$
\end{minipage} 
& 
Non-overlapping Faint and Bright.
\\
\tblmidlineb 
\gktwo:All
& 
F:All + B:All + $\left<\delta_B \delta_F\right>$ 
&
\begin{minipage}[t]{\mywidth}
FxB-All removing $\left<\delta_B \gamma_F\right>$
\end{minipage} \\
\tblspaceb
\gkone:All
&
B:All + $\left<\delta_F \delta_F\right>$ + $\left<\gamma_F \gamma_F\right>$ +
     $\left<\delta_F \delta_B\right>$ + $\left<\delta_B \gamma_F\right>$
&
\begin{minipage}[t]{\mywidth}
FxB-All removing $\left<\delta_F \gamma_F\right>$
\end{minipage} \\
\tblspaceb
\gkthree:All
&
B:Counts + F:Counts + $\left<\gamma_B \gamma_B\right>$ + $\left<\gamma_F \gamma_F\right>$
&
\begin{minipage}[t]{\mywidth}
FxB-All. removing all cross-correlation of  counts and shear.\end{minipage} \\
\tblmidlineb
\gkfb:Counts
&
$\left<\delta_B \delta_B\right>$ + $\left<\delta_F \delta_F\right>$
&
\begin{minipage}[t]{\mywidth}
FxB-Counts removing the Bright-Faint cross-correlations of counts.
\end{minipage} \\
\tblspacebb
\gkfb:All
&
F:All + B:All
&
\begin{minipage}[t]{\mywidth}
FxB-All removing all the Bright-Faint cross-correlations, i.e.
$\left<\delta_F \delta_B\right>$ and $\left<\delta_B \gamma_F\right>$.
Equivalent to F+B:All with covariance between F and B.
\end{minipage} \\
\tblpadb
\hline
\end{tabular}
\end{center}
\caption{Table summarizing the probe combinations and the correlations
included. The first column is the notation, the second column give the
correlations included and the third column is a short description. Each row
correspond to a different probe combinations. The first block show standard
combinations, the second block show combinations removing counts-shear
correlations and the third block includes probes without cross-correlations
of the two samples. Here "Counts" only include galaxy counts, while "All" also
includes shear. The observables $\left<\gamma_B \gamma_F\right>$ and $\left<\delta_F
\gamma_B\right>$ are also included, but not listed as their contributions are
minor.}
\label{notation_one}
\end{table*}

%% file: results.tex
\newcommand{\tblskip}{\noalign{\vskip 1.0mm}}
\newcommand{\tblpad}{\noalign{\vskip 1mm}}
\newcommand{\tblmidline}{\tblskip\hdashline\tblskip}

In this section we investigate the combined constraint from galaxy counts and
shear, treating each survey as a separate galaxy population. The first five
subsections compare how different effect as RSD, BAO, lensing and intrinsic
correlations contribute to the forecast. In the last subsection we present the
main forecast table and discuss the relative contribution of each effect and
the impact with overlapping photometric and spectroscopic surveys. Last we look
at the effect of magnification. Contour plots can be found in appendix
\ref{app_contours}.

\subsection{Auto versus cross-correlations}
\label{subsec:res_radial}

This subsection study how the galaxy clustering and lensing observables affects
the forecast. The significant correlations for close redshift bins are the
shear-shear $\left<\gamma\gamma\right>$ and counts-counts
$\left<\delta\delta\right>$ correlations. Here the counts-counts correlation of
the spectroscopic sample include a strong RSD signal and intrinsic
cross-correlation between nearby redshift bins. For large redshift bin
separation the counts-shear $\left<\delta\gamma\right>$ cross-correlations are
the strongest. To separate the contribution from different cross-correlations,
we introduce the variable $\dzmax$. All correlations $C_{ij}$ are required to
satisfy

\be
\left|z_j - z_i\right| \leq \dzmax
\ee

\noindent
where $z_i,z_j$ respectively are the mean of redshift bin $i$ and $j$. This
requirement only applies when specified for the figures. Which
cross-correlations enters is discussed together with the forecast results in
the next paragraphs.

\xbf
\xfigure{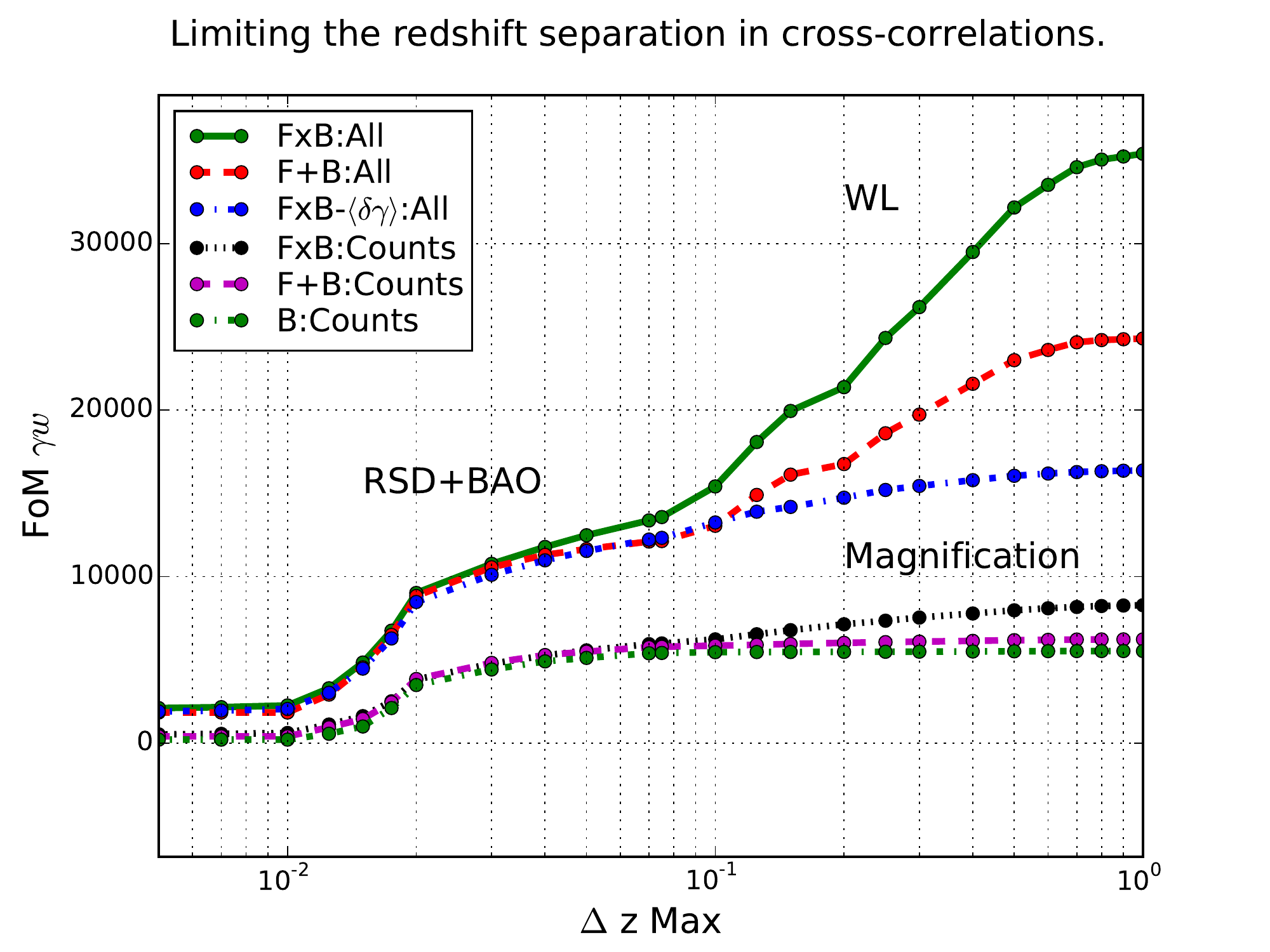}
\xfigure{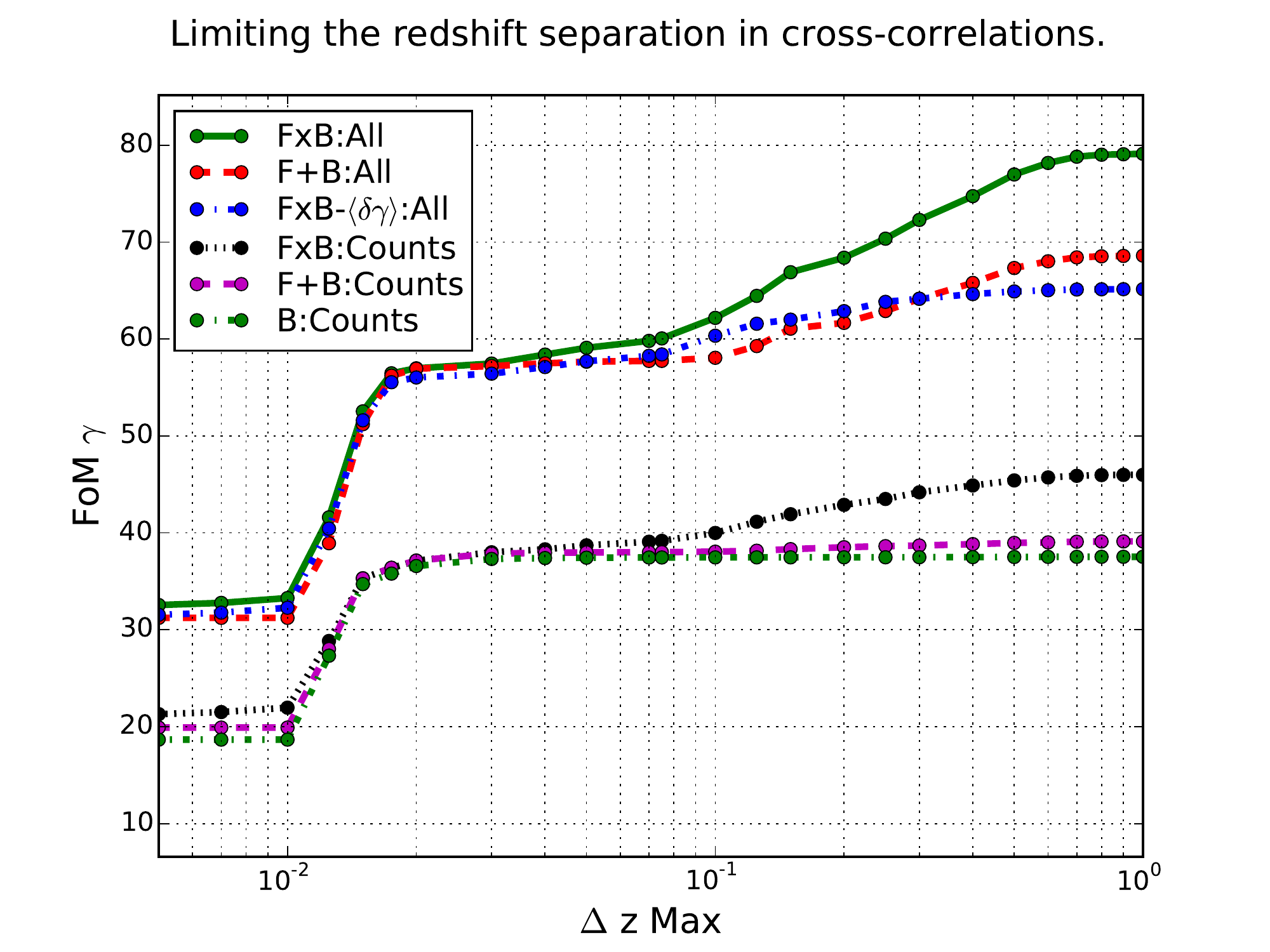}
\caption{The FoM dependence on the maximum redshift separation for
cross-correlations included in the forecast ($\dzmax$). Lines corresponds to
the probes FxB:All, F+B:All, \gkthreens:All, FxB:Counts, F+B:Counts and
B:Counts. The top and bottom panel respectively corresponds to $\fomc$ an
$\fomg$.}

\label{res_comb_overall}
\xef

The auto-correlations are alway included from the $\dzmax$ definition, which
for probes with lensing includes the shear-shear auto-correlations. In Fig.
\ref{res_comb_overall} this shows at $\dzmax = 0$ as a gap between lines which
includes shear (All) or not (Counts). The lensing with its broad kernel can be
seen to better measure dark energy (top panel, also includes $\gamma$) than the
growth of structure (bottom panel). Overall, galaxy shear lead to 4-5 times
improvement for the combined figure of merit ($\fomc$). In the region
$0.01<\dzmax<0.1$ the forecast also include cross-correlations
between Bright/spectroscopic redshift bins, with a significant jump when
including the cross-correlation with the adjacent bin. These cross-correlations
contribute significantly and are studied in later subsections in the context of
the Limber approximation (\ref{subsec:res_limber}), RSD (\ref{subsec:res_rsd})
and BAO (\ref{subsec:res_bao}).

The \gkthree\ lines are the forecast of FxB:All without count-shear
correlations. At larger $\dzmax$ the counts-shear lensing becomes important,
which can be seen from FxB:All and F+B:All having higher $\fomc$ than
\gkthree. Note that higher $\dzmax$ also includes shear-shear tomography,
which also enters in \gkthree. Another lensing effect is the magnification of
the galaxy counts (paper-I). Intrinsic clustering dominated the counts-counts
signal at low redshift separation, while magnification only becomes important
at higher separations where the intrinsic correlation also vanish. The
separation of FxB:Counts and F+B:Counts at high $\dzmax$ is due to
magnification. Magnification is also included in F and B, but the impact is
strongest for the combined overlapping samples. In subsection
\ref{subsec:res_magn}, we study the effect of magnification when the surveys
both include galaxy counts and shear.

\xbf
\xfigure{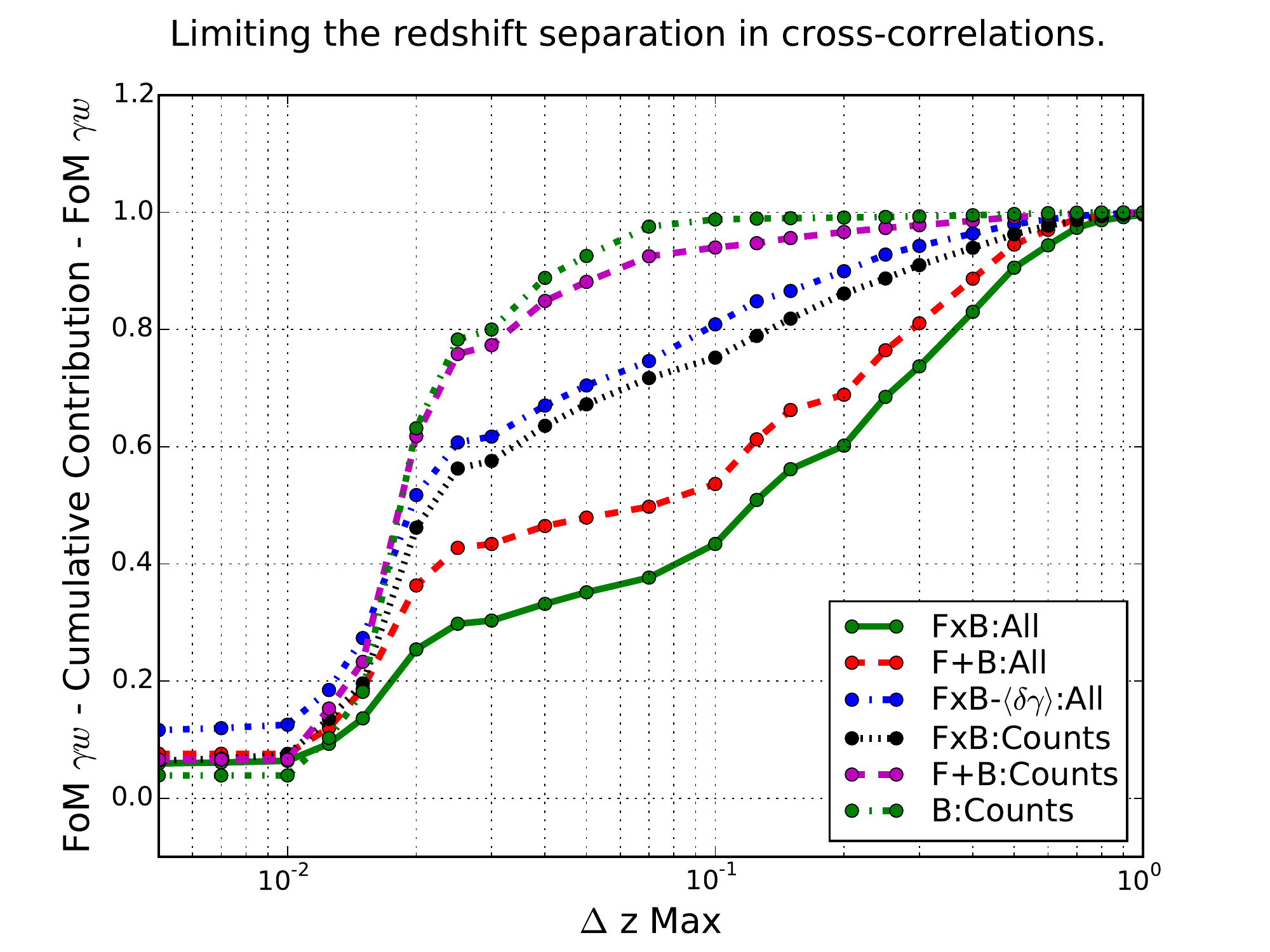}
\xfigure{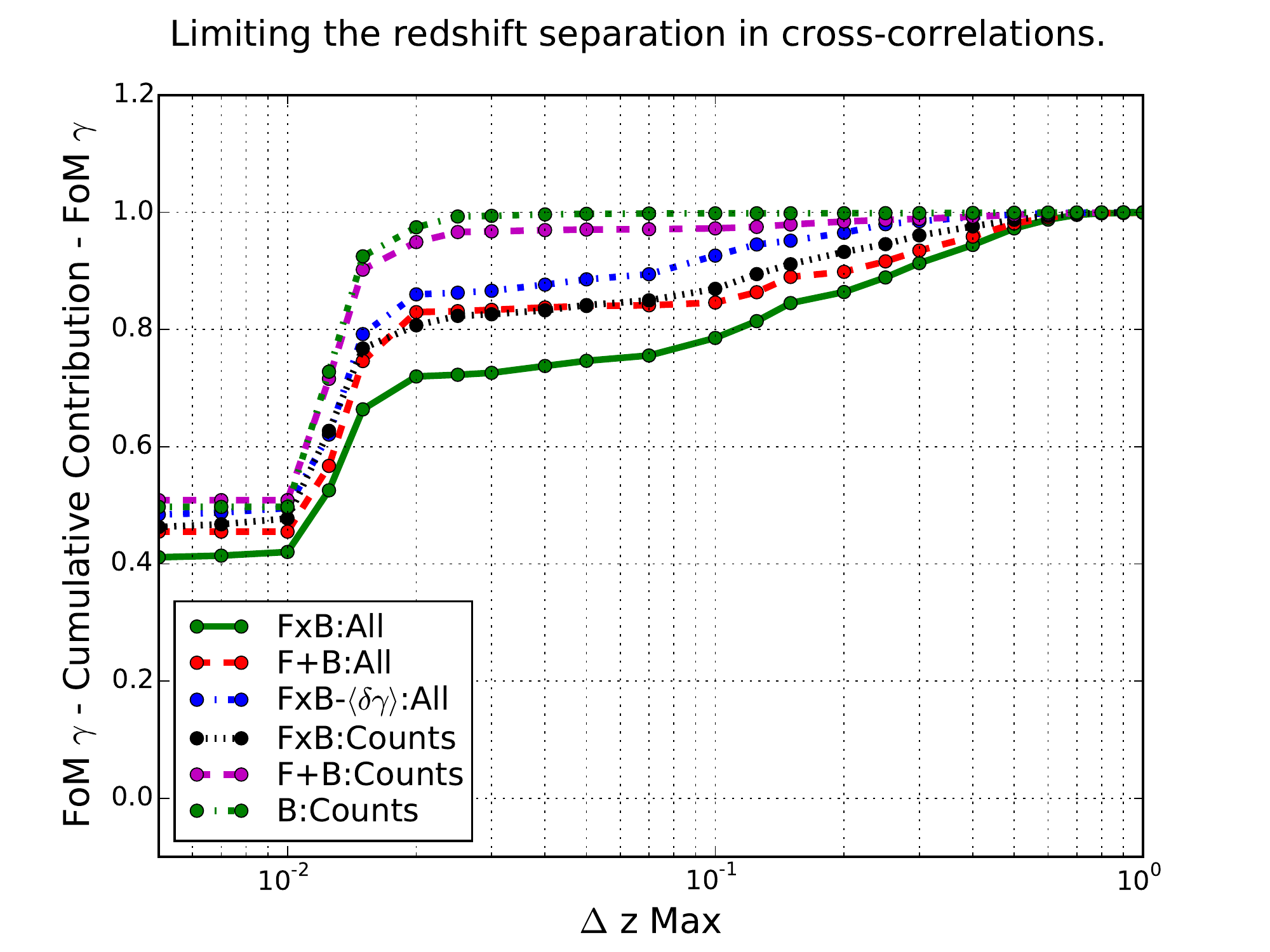}
\caption{Normalized cumulative FoM contributions for different probes. The
x-axis is $\dzmax$ and the FoMs are normalized by $\text{FoM}(\dzmax=1)=1$.
Each line is a probe and the two panels show the result for different
FoMs.}
\label{res_comb_ratio}
\xef

Fig.\ref{res_comb_ratio} shows the FoM normalized to $\text{FoM}(\dzmax=1) =
1$ where the gain has saturated. The same information is already presented
(Fig. \ref{res_comb_overall}), but these plots are better to discuss the
relative contribution of different correlations. In $\fomc$ (top panel) there
is large spread between the lines. For B:Counts mostly close ($\dzmax < 0.1$)
correlations are important. When including magnification (FxB:Counts) or the
shear-shear tomography (\gkthree), there is some more benefit from
cross-correlations of widely separated redshift bins. The bottom lines are
FxB:All and F+B:All where counts-shear contribute significantly and many
different correlation types contribute to the constraints. There is much less
difference for $\fomg$ (bottom panel), where the auto-correlations account for
40-50\% for all probes. For B:Counts the intrinsic counts-counts correlation
between bins provide the rest, while FxB:All has a 25\% contribution from
counts-shear lensing.

\xbf
\xfigure{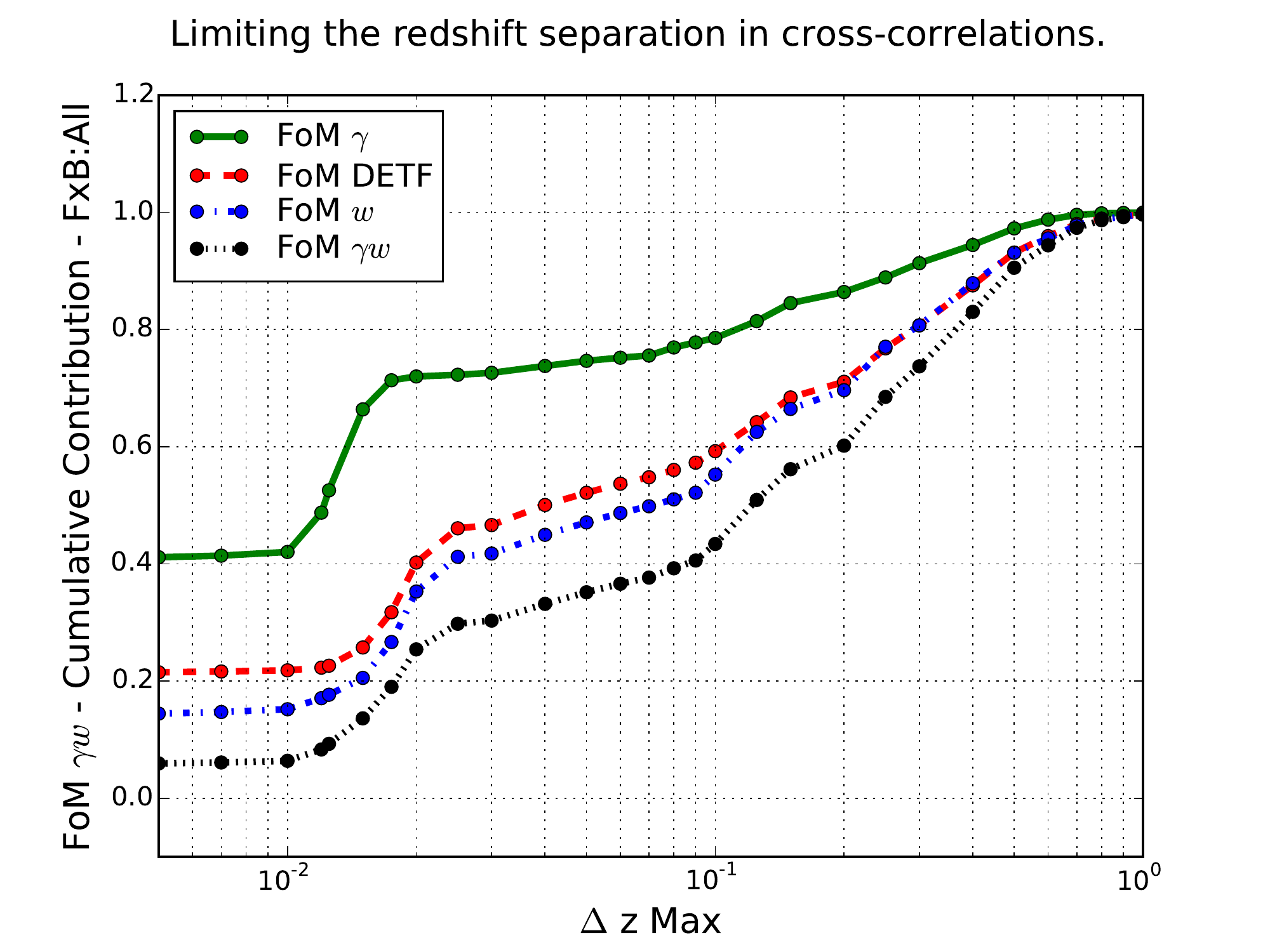}
\caption{Normalized cumulative FoM contributions for different probes. The
x-axis is $\dzmax$ and the FoMs are normalized by $\text{FoM}(\dzmax=1)=1$. All
results are for FxB:All and the lines corresponds to $\fomg$, $\fomdetf$,
$\fom$ and $\fomc$.}
\label{res_comb_foms}
\xef

Fig.\ref{res_comb_foms} is similar to Fig. \ref{res_comb_ratio}, but compares
the normalized cumulative constraints of FxB:All for the different FoMs:
$\fomg$, $\fomdetf$, $\fom$ and $\fomc$ in the same plot. The $\fomg$ line
depends strongest on the auto-correlation. This is expected as the galaxy
clustering (counts-counts) is important for measuring the growth. Interestingly
the next two lines are $\fomdetf$ and $\fom$, while $\fomc$ which includes both
dark energy and the growth ($\gamma$) benefit the most from different
correlations. Also, fixing the bias change which correlations that contribute
(plot not shown), while keeping the FoM/line order. How marginalising over the
bias change the forecast is an important part of this paper and is studied
further in paper-III.

\subsection{Limber approximation}
\label{subsec:res_limber}
\xbf
\xfigure{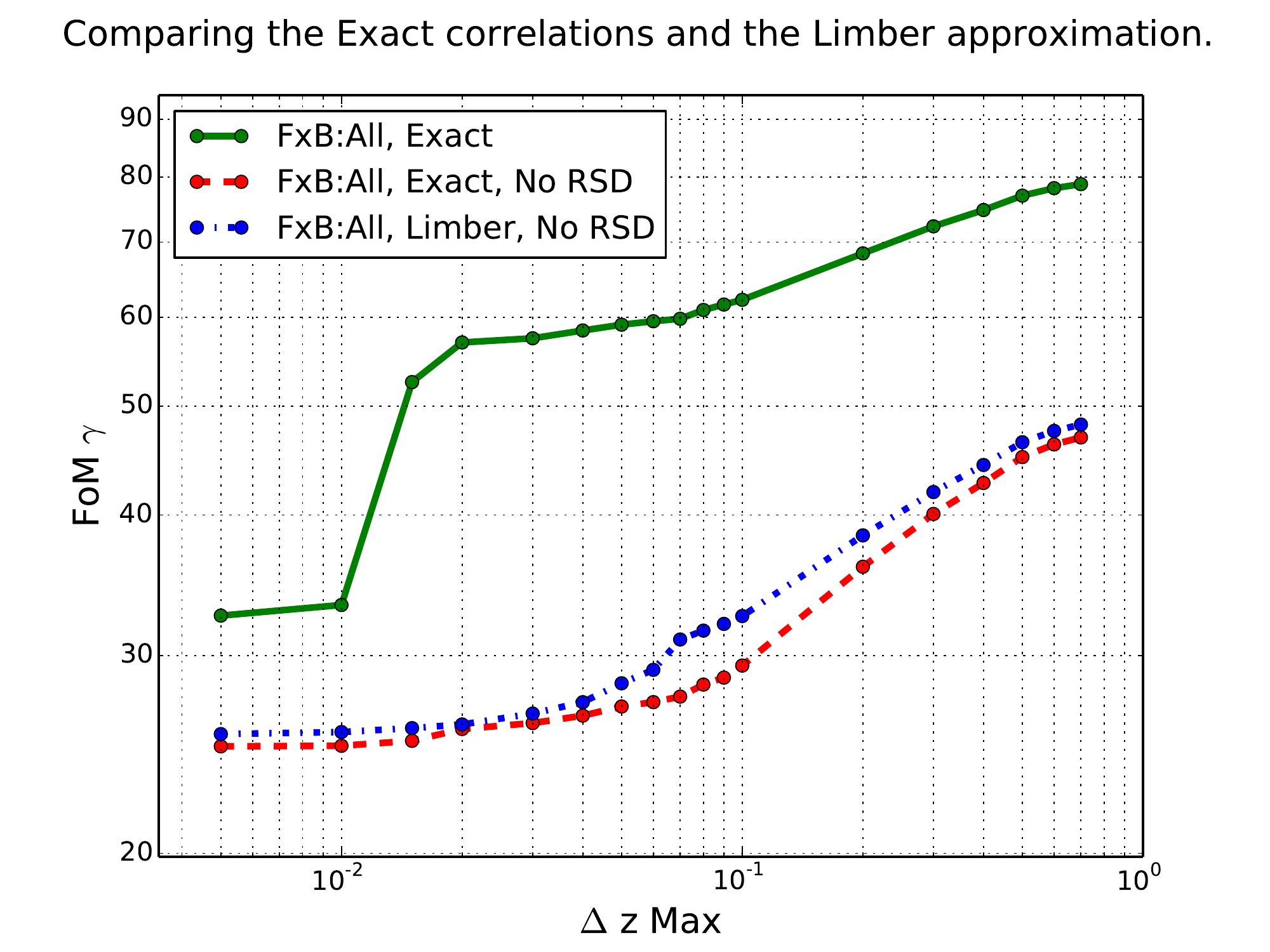}
\xfigure{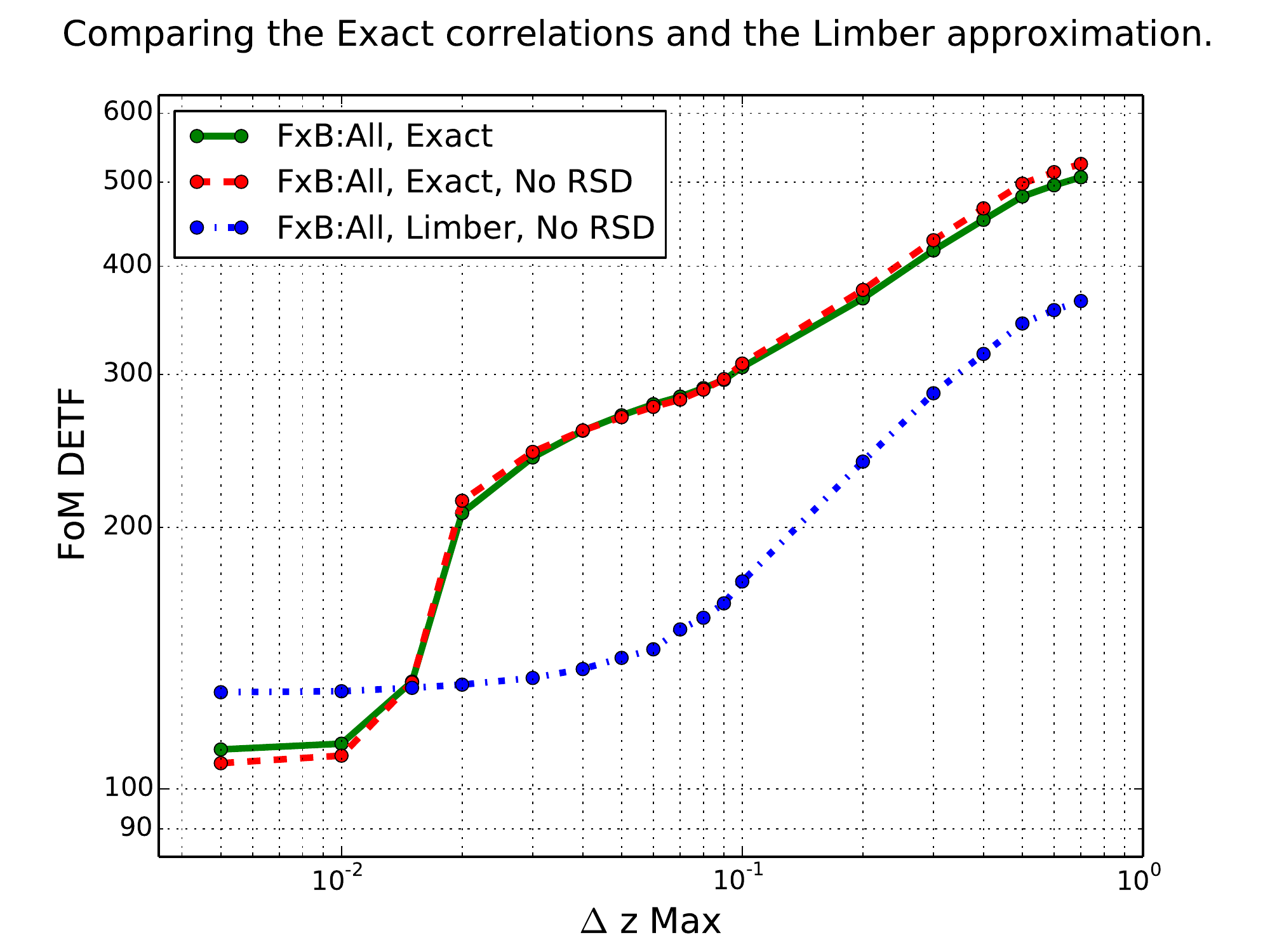}
\caption{Forecast for FxB:All using the exact calculations and the Limber
approximation. The first two lines respectively show the FoM in redshift and
real space, while the last use the Limber approximation in real space. The top
and bottom panel respectively show $\fomg$ and $\fomdetf$.}
\label{res_limber_exact}
\xef

Paper-I compared the correlations estimated using the exact calculations to the
correlations when using  the Limber approximation. For narrow redshift bins of
$\dzbin = 0.01(1+z)$, the Limber approximation can overestimate the galaxy
counts auto-correlations by a factor of 2-3. Further, in the Limber
approximation there is no counts-counts cross-correlations between
non-overlapping redshift bins, which is not a good approximation for $\dzbin=
0.01(1+z)$ wide bins in the Bright sample.

Fig.\ref{res_limber_exact} compare the exact calculations with the Limber
approximation. Included in the panels is one line showing the exact
calculations with RSD, while the other two lines are the exact calculation and
Limber approximation in real space (No RSD). The redshift space distortion
signal in the correlation is powerful, especially in measuring $\gamma$.
Comparing the three lines show how cross-correlations and redshift space
distortions contribute to measuring dark energy and the growth of structure.
For $\gamma$ including the cross-correlations has little effect, while the
redshift space distortions improve $\fomg$ for FxB:All by 70\%. On the other
hand, for $\fomdetf$ the cross-correlations of galaxy count in the radial
direction is powerful, while the RSD signal contributes little. One can
understand the main traits from the amplitudes and shapes of the correlations.
The $\gamma$ parameter changes the clustering amplitude, while dark energy
parameters $\omega$ more directly affects the shape.

For $\fomdetf$ without RSD (real space) the exact calculations and Limber
approximation results cross around $\dzmax = 0.015$. The width of the
spectroscopic redshift bins here is $\dzbin = 0.01 (1+z)$ and around the
crossing the exact calculations begin to include correlations with nearby
redshift bins. These are important for dark energy constraints (subsection 
\ref{subsec:res_bao}). Also, similar to $\fomg$, when counts-shear becomes
important at large $\dzmax$, the difference decreases because of the smaller
error in Limber approximation.

The higher galaxy counts auto-correlations in the Limber approximation reduce
the impact of shot-noise. One can in Fig. \ref{res_limber_exact} see how the
$\fomg$ line in top panel of Fig.\ref{res_limber_exact} is slightly lower for
the Exact calculation than the Limber approximation. For larger $\dzmax$, the
lines first diverge before converging when also including counts-shear
cross-correlations (high $\dzmax$). While the Limber approximation is accurate
for the counts-shear signal, the higher galaxy counts lead to an overestimated
error (see Eq. \ref{covariance}). As a result, the counts-shear correlations
contribute less in the Limber approximation.

\subsection{Resolution in redshift}
\label{subsec:res_resolution}

\begin{figure}
\xfigure{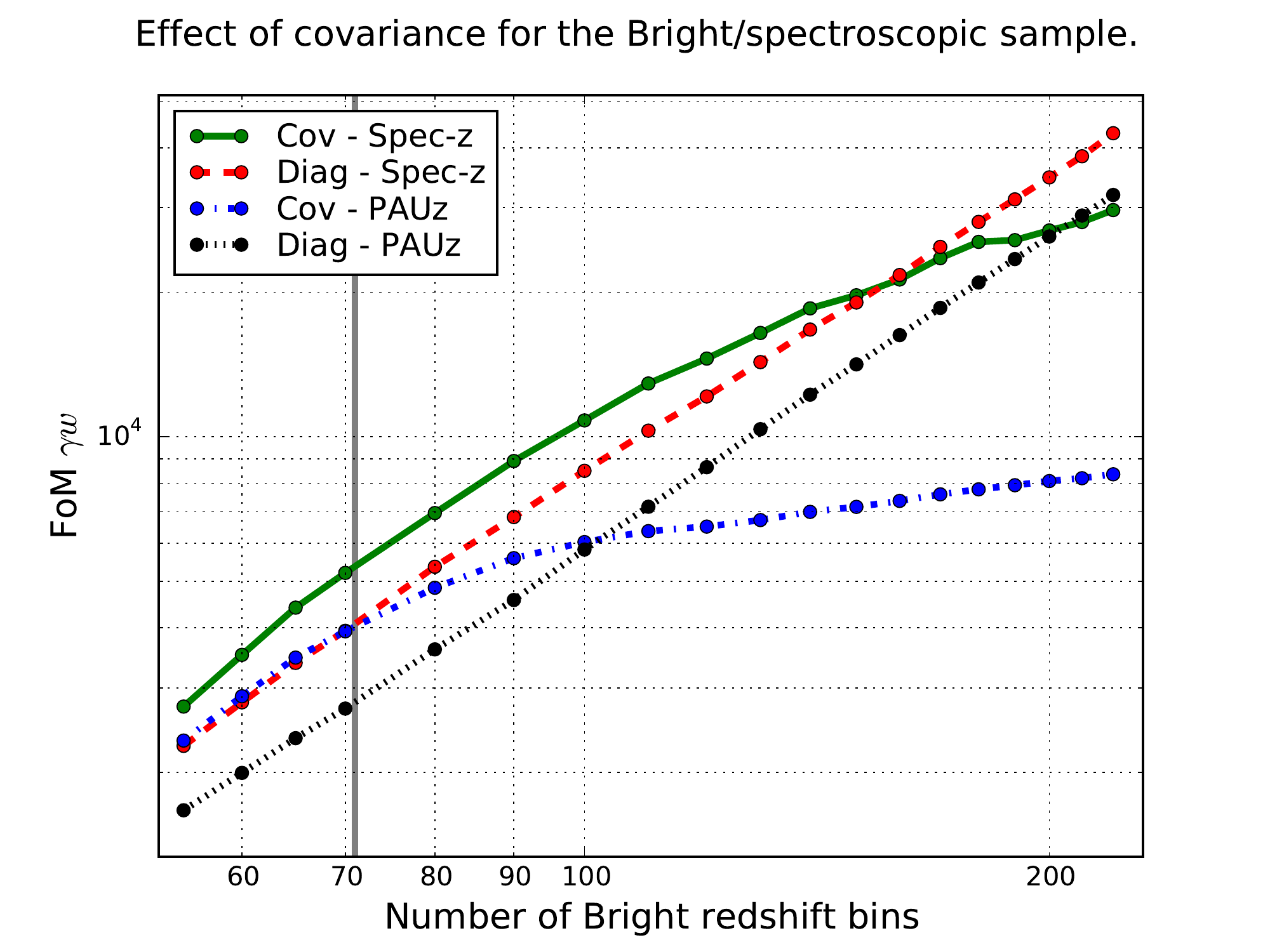}
\caption{Effect of the covariance and photo-z. The lines all show $\fomc$ for
galaxy counts for the spectroscopic sample, with the number of bins varying on
the x-axis. Two lines (cov) include the full covariance, while the other two
(only diag) only use the variance. Each of these configuration is run either
using spectroscopic redshifts (spec-z) or narrow band photometry (PAUz,
$\sigma_{68}/(1+z) = 0.0035$)). A vertical line mark the fiducial number of
spectroscopic redshift bins (\nbright).}
\label{resol_bright}
\end{figure}

\newcommand{\nbintext}{
Let $z_i$ denote the edges between redshift bins and where $z_0$ is the start
of the redshift range. A frequently used redshift binning is $z_n = z_{n-1} +
(1+z_{n-1})*w$ where the constant $w$ give the bin width. Provable by
mathematical induction, then

\be
z_n = (1+z_0){(1+w)}^n - 1.
\ee

\noindent
is the nth edge between the redshift bin. For a binning $\Delta_z = w (1+z)$, then

\be
w = \sqrt[\leftroot{4}\uproot{6}N]{\frac{1+z_{\text{Max}}}{1+z_0}} - 1
\ee

\noindent
divide the interval $[z_0, z_{\text{Max}}]$ in n bins into N redshift bins.}

Increasing the number of spectroscopic bins result in better constraints. The
redshift bin width in a 2D forecast corresponds approximately to the maximum
scale $k_{max} = 2 \pi / \lambda_{min}$, where $\lambda_{min}$ is the comoving
width of the redshift bins \citeind{asorey}. The gain for small enough bins
when increasing the number of bins therefore mainly come from probing smaller
scales. Note that having such a large number of bins can lead to include more
nonlinear modes in the radial direction than in the angular direction. To be
consistent we need to limit the number of radial bins to the corresponding lmax
scale (see \citedir{asorey,asorey2}). In our case Nz~70 is the corresponding
number and in this regime PAUz is quite close to Spec-z. In this section we
are not comparing to a 3D forecast, but focus on the effect of the covariance
between the observables.

Fig. \ref{resol_bright} show $\fomc$ for an increasing number of bins, where
the bin width $\Delta z = w (1+z)$ is set by the number of redshift
bins\footnote{\nbintext}. Focusing first on the result for many redshift bins,
one expect the covariance to be important for increasingly thinner bins. The
galaxy density per bin also decrease, but the fiducial sample is dense (0.4
gal/sq.arcmin) and the effect of shot-noise is less important (plot not shown).
Assume the auto-correlations are close to equal in two bins ($C_{AA} \approx C_{BB}$) and define
$\alpha \equiv C_{AB}/C_{AA}$. The Pearson correlations, $R[A,B] =
\text{Cov}(A,B)/\sqrt{\text{Var}[A]*\text{Var}[B])}$, for the covariance matrix
(Eq.\ref{covariance}) is then

\begin{align}
R[\text{Auto(AA)}, \text{Auto(BB)}]  &\approx \alpha^2 \\
R[\text{Auto(AA)}, \text{Cross(AB)}] &\approx \alpha \sqrt{2} /\sqrt{1 + \alpha}
\label{covexample}
\end{align}

\noindent
when ignoring the shot-noise. From these equation, the covariance increase for
thinner bins which has higher $\alpha$ (see paper-I). The largest covariance is
not between auto-correlations, but between the auto and cross-correlations.
Previous studies expected the covariance to saturate the result, but could not
demonstrate this due to technical difficulties with many bins \footnote{In this
subsection the forecast exclude all counts-counts cross-correlations $C_{ij}$
with $0.1 < |z_j - z_i|$ for redshift bins $i,j$. These include little
cosmological information (see magnification subsection \ref{subsec:res_magn})
and removing these help to reduce the dimension of the covariance matrix.}
\citeind{asorey,classgal}. In Fig. \ref{resol_bright} the covariance limits the
results, with the lines being even flatter (and numerical unstable) when
approaching 300 bins. The forecast also saturate for $\fom$, $\fomg$ and
$\fomdetf$ (not shown). For PAUz the forecast FoMs become flat earlier (less
bins), since the photo-z also correlate the fluctuations in the different
redshift bins.

For an intermediate low number of bins (50-100), the covariance between
observables increase $\fomc$. This result is counter-intuitive, but is similar
to the sample variance cancellations for multiple galaxy tracers.  When two
observables depend differently on nuisance parameters (e.g. bias), the
covariance between the observables introduce a covariance between the nuisance
parameters. The additional covariance between the bias parameters reduce their
freedom, which increase cosmological constraints when marginalised over.  The
covariance naturally also reduce the information, since the observables are no
longer independent. If the forecast improve or degrade depends on the details
of these competing effects (see \citedir{sameskyX}).

\begin{figure}
\xfigure{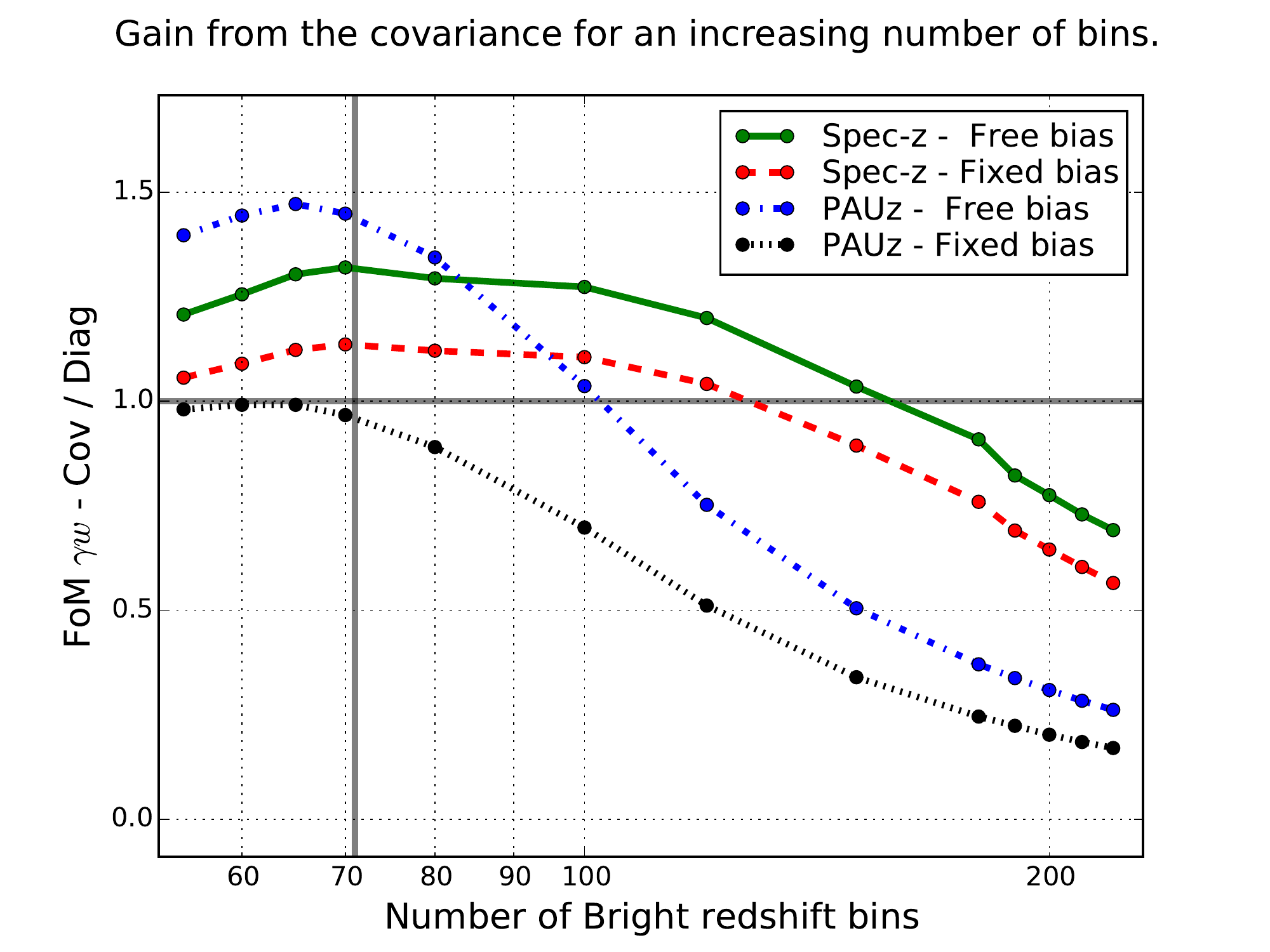}
\caption{The gain from the covariance. The figure show the forecast ratio
between including the full covariance or only the diagonal entries. All lines
show $\fomc$ for the Bright sample using galaxy counts (B-Counts) as a function
of redshift bins. Two lines use a spectroscopic sample (spec-z), while the
other two use narrow-band photo-z sample (PAUz). For two lines we marginalise
over the bias (free bias), while for the other two the bias is fixed (fixed
bias). A vertical line mark the fiducial number of spectroscopic redshift bins
(\nbright).}
\label{radcov_gain}
\end{figure}

Fig. \ref{radcov_gain} show the $\fomc$ ratio between including the full
covariance or only the diagonal entries (variance). For spectroscopic redshifts
and marginalising over the bias (free bias), the covariance increase $\fomc$
until about 160 spectroscopic bins. This show that the covariance between
different redshift slices increase constraints through reduced the sample
variance. When fixing the bias the gain is about 3 times lower, but it is still
a 10\% effect. We attribute this to a changed covariance between the
cosmological parameters, some of which we marginalise over. While the
constraints can be higher for only the variance, the covariance should be
included in parameter fits to not bias the results. With narrow band photo-z
(PAUz), the effect of the covariance change. For a free bias the gain is
higher, while there is no benefit when fixing the bias.

\begin{figure}
\xfigure{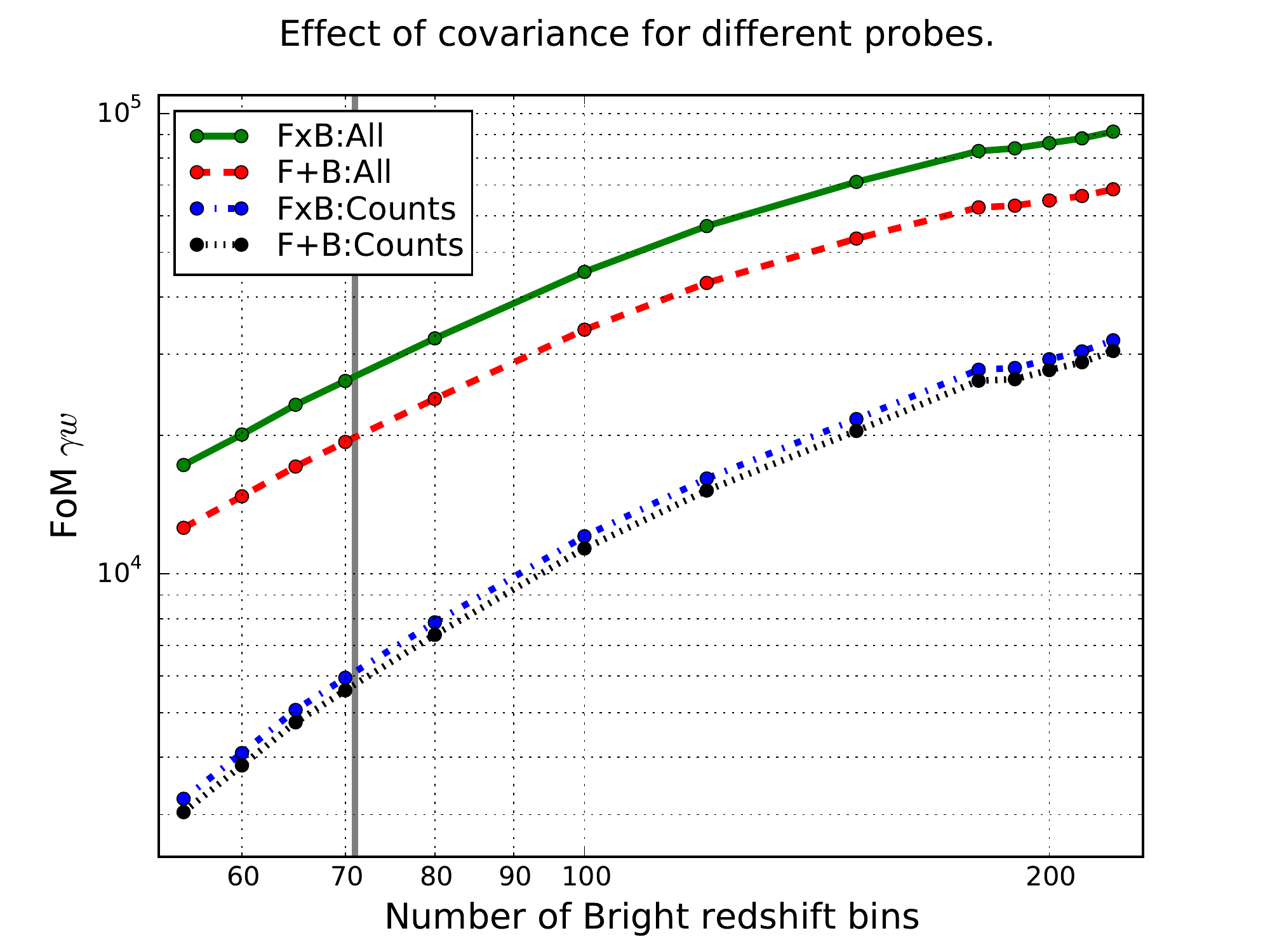}
\caption{The forecast for different probes and increasing number of bins. On
the x-axis is the number of spectroscopic redshift bins, while the lines show
$\fomc$ for the probe combinations FxB:All, F+B:All, FxB:Counts and
F+B:Counts. A vertical line mark the fiducial number of spectroscopic redshift
bins (\nbright).}
\label{resol_all}
\end{figure}

How does the same-sky (FxB/F+B) conclusions depend on the number of
spectroscopic bins? Fig. \ref{resol_all} show the combined forecast (FxB, F+B)
for the fiducial case, which use spectroscopic redshift and include the
covariance. The Bright sample (Fig. \ref{resol_all}) benefit from
more redshift bins. This however also increase the number of counts-counts
cross-correlations with the photometric sample and cross-correlations of
spectroscopic galaxy counts with shear. For both "All" and "Counts" the FxB/F+B
ratio is nearly constant over a wide number of spectroscopic redshift bins. This
shows our conclusions on the same-sky issue is quite robust with respect to the
number of redshift bins.

\subsection{Redshift space distortions (RSD)}
\label{subsec:res_rsd}

\newcommand{\galP}{\tilde{P}_{\text{Gal}}}
Redshift space distortions affects the overdensities of galaxies. A matter
overdensity attract galaxies, which change their velocities and introduce a
change in redshift. At linear level, the change in galaxy count overdensities
is the Kaiser effect. The redshift space galaxy spectrum $\galP(k)$ is then

\be
\galP(k, \mu) = {\left(b + \mu^2 f\right)}^2 P(k)
\ee

\noindent
where $P(k)$ is the real space matter power spectrum, $b$ is the galaxy bias,
$\mu$ is the cosine of line of sight angle and $f \equiv \Omega_m(z)^\gamma$.
In the forecast, the RSD effect enters in the 2D correlations. Overdense
regions attract nearby galaxies, which can move galaxies between redshift bins.
This effect often increase the amplitude of the 2D-correlations (see paper-I).

\xbf
\xfigure{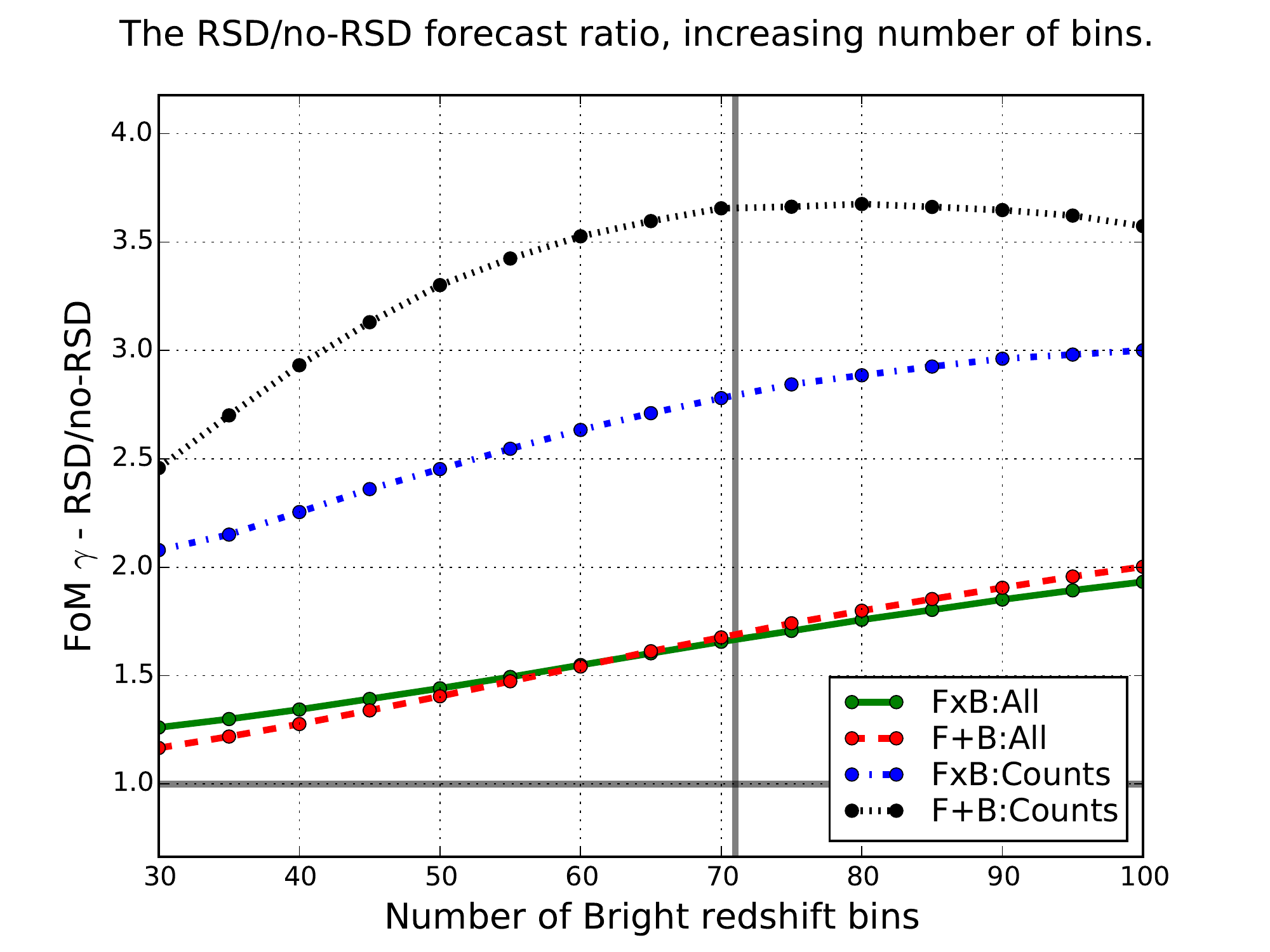}
\caption{The $\fomg$ ratio between a redshift and real space forecast when
varying the number of spectroscopic redshift bins. A vertical line mark the
fiducial number of spectroscopic redshift bins (\nbright). The four lines corresponds
to FxB:All, F+B:All, FxB:Counts and F+B:Counts.}
\label{res_rsd_nbins}
\xef

The redshift space distortions is a powerful effect for measuring $\gamma$.
Fig. \ref{res_rsd_nbins} show how including RSD in the correlations improves
$\fomg$ by a factor between 1.5 and a few. For the other FoMs (not shown) RSD
decrease the result with 0-5\%. Now we first focus on the results for counts.
Observing galaxy counts over separate skies (F+B:Counts) is the combination
which benefits most from RSD. For only B:Counts (not shown), the RSD improves
for the fiducial binning the constraints with a factor of 3.9. The added RSD
component is independent of bias, and therefore reduce the degeneracies between
$\gamma$ and the bias. In FxB:Counts, the surveys are overlapping and the
samples (F and B) can be cross-correlated. The cross-correlations and also
sample variance cancellation directly from overlapping volumes (subsection
\ref{subsec:res_gain}) improve bias constraints and RSD is therefore less
important for FxB:Counts.

One should note, comparing models with and without RSD in the angular
correlations is slightly misleading. The forecast in redshift space or real
space includes the same correlations and only differ by including the RSD
component in the correlations. One can therefore not assume RSD always improves
the parameter constraint, but the benefit depends largely on the resulting
correlations between the parameters. As seen in last paragraph, the measurement
of $\gamma$ improves greatly from RSD. On the other hand, in the theoretical
real space angular correlations, there is radial information in the
cross-correlations between redshift bins. Including RSD will, as we will see,
reduce the dark energy constraints from intrinsic galaxy counts
cross-correlations with nearby redshift bins.

When including shear the importance of RSD naturally decrease
(Fig.\ref{res_rsd_nbins}, FxB:All and F+B:All). These probes also include the
shear-shear signal, which is unaffected by the RSD. Even if the ratios are
lower, the factor of 2 is still a good improvement. Since F+B:Counts benefit
more than FxB:Counts from RSD, the separation between F+B:All and FxB:All is
smaller than expected. One can understand this from looking at the counts-shear
variance. The variance for the cross-correlation of a foreground galaxy
counts ($\delta_i$) density with background shear ($\gamma$) is (see Eq.
\ref{covariance})

\be
\text{Var}(\left<\delta_i \gamma\right>) =  N^{-1}(l) [\left<\delta_i
\delta_i\right> \left<\gamma \gamma\right> + \left<\delta_i \gamma\right>^2]
\label{cgk_suppress}
\ee

\noindent
where $N(l)$ is the number of modes. The $\left<\delta_i \gamma\right>$ signal
and the second term in the variance is approximately independent of redshift
space distortions. On the other hand, the $\left<\delta_i \delta_i\right>$
auto-correlations increase strongly from RSD. Since the error increases,
including RSD in the forecast reduce the importance of counts-shear.

\xbf
\xfigure{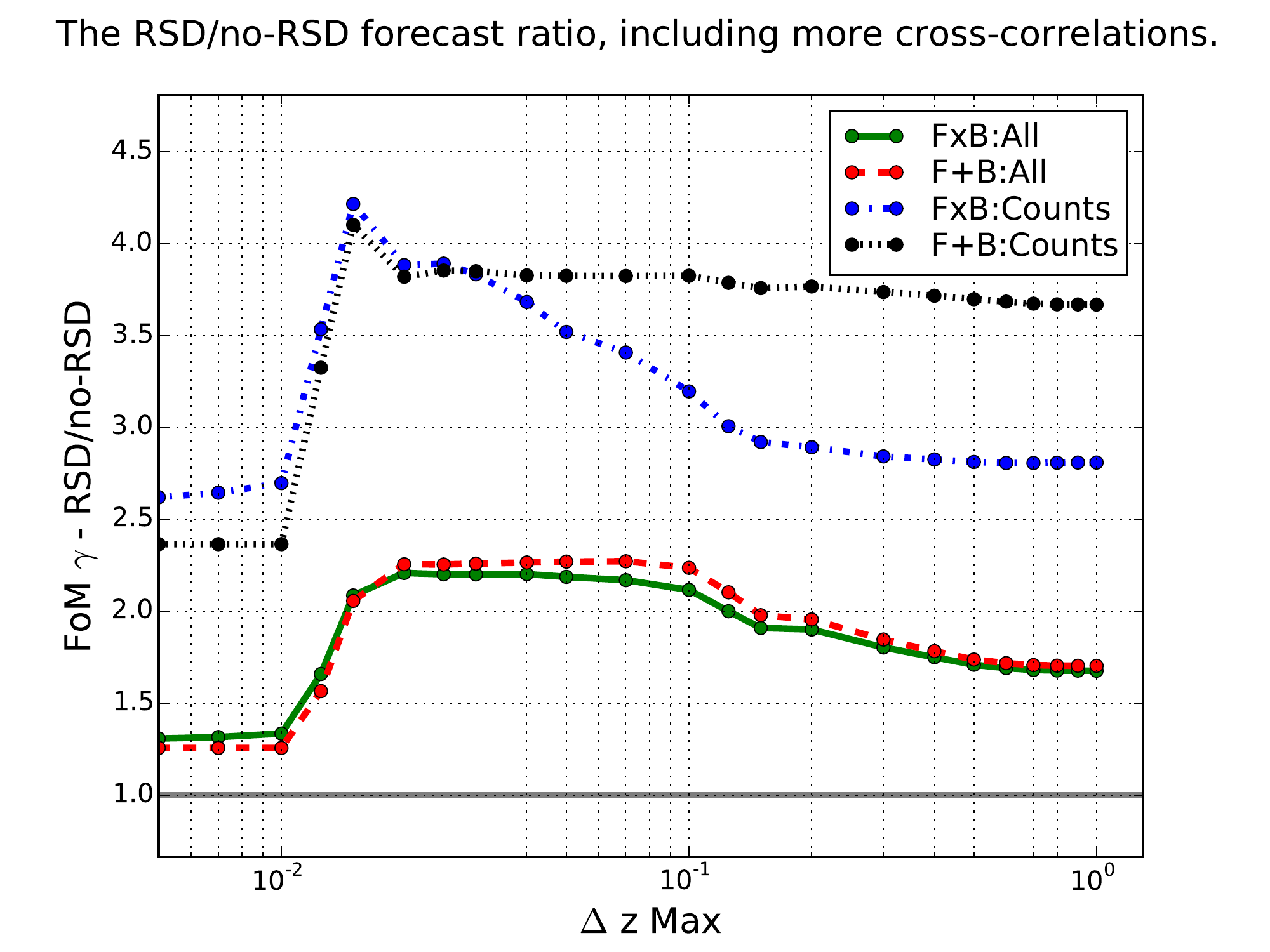}
\xfigure{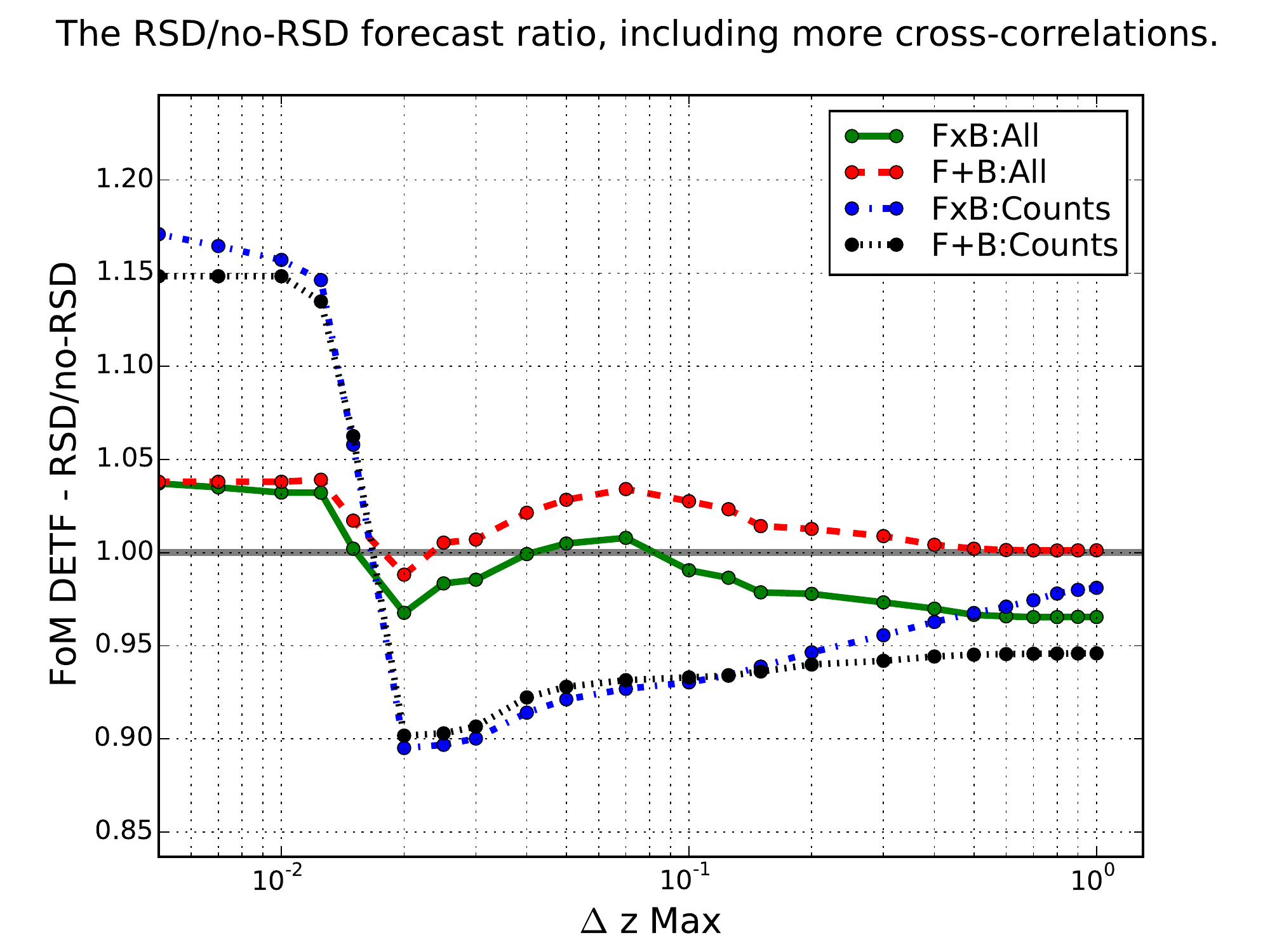}
\caption{FoM ratio between redshift and real space when varying the largest
redshift separations in the cross-correlations ($\dzmax$). In the two plots
corresponding to $\fomg$ and $\fomdetf$ the four lines are the probes FxB:All,
F+B:All, FxB:Counts and F+B:Counts.}
\label{res_rsd_radial}
\xef

Top panel of Fig.\ref{res_rsd_radial} shows how including RSD affects the
forecasts. Instead of studying the FoMs as a function of number of redshift
bins, this figure use the fiducial binning and vary $\dzmax$. Subsection
\ref{subsec:res_radial} explained how $\dzmax$ can be used to distinguish
between contributions from auto-correlations, cross-correlation with nearby
redshift and counts-shear weak lensing. The largest RSD effect is for $\fomg$
and F+B:Counts increase by a factor of 3.6 with respect to the real space
forecast. When including cross-correlations between nearby bins, the importance
of RSD increase for all probe combinations. In paper-I we showed how RSD affect
the auto and cross-correlations differently, which here improve the growth
constraints. For FxB:Counts, higher $\dzmax$ also include weak lensing
magnification. Magnification adds an additional bias measurement, therefore
decreasing the impact of redshift space distortions.

Last, the $\fomdetf$ (Fig.\ref{res_rsd_radial}, bottom panel) include some
interesting trends. For the auto-correlation the RSD improve the dark energy
constraints. Around $0.01 < \dzmax < 0.02$ the forecast also include galaxy
counts cross-correlations between spectroscopic redshift bins. Then the
$\fomdetf$ ratio suddenly drops because the RSD suppress the cross-correlation
with nearby redshift bins, which are important for dark energy constraints.
Since FxB:Counts depend stronger than F+B:Counts on magnification, the lines
separate at high $\dzmax$. The magnification signal, which has similar
covariance to counts-shear and benefit from RSD reducing amplitude of the
auto-correlations. Since FxB:Counts depends stronger on magnification than
F+B:Counts, the lines separate at high $\dzmax$.

\subsection{Baryon Acoustic Oscillation (BAO)}
\label{subsec:res_bao}
\newcommand{\kpar}{r_{\parallel}} \newcommand{\kperp}{r_\perp}

In the observed galaxy distribution BAO is a characteristic scale ($\sim150$ Mpc
today), which is measurable both in the transverse/angular and radial/redshift
direction \citeind{cabreplot, lssgamma}. Observing the BAO can therefore probe cosmology
by measuring the comoving and angular diameter distance. Accurately predicting
the power spectrum is done by solving the Boltzmann equation. The Eisenstein-Hu
analytical power spectrum formula we use here is less accurate, but can be
estimated the power spectrum both only with the continuum and including the BAO
wiggles. In this subsection we compare the forecasts with and without the BAO
feature.

\xbf
\xfigure{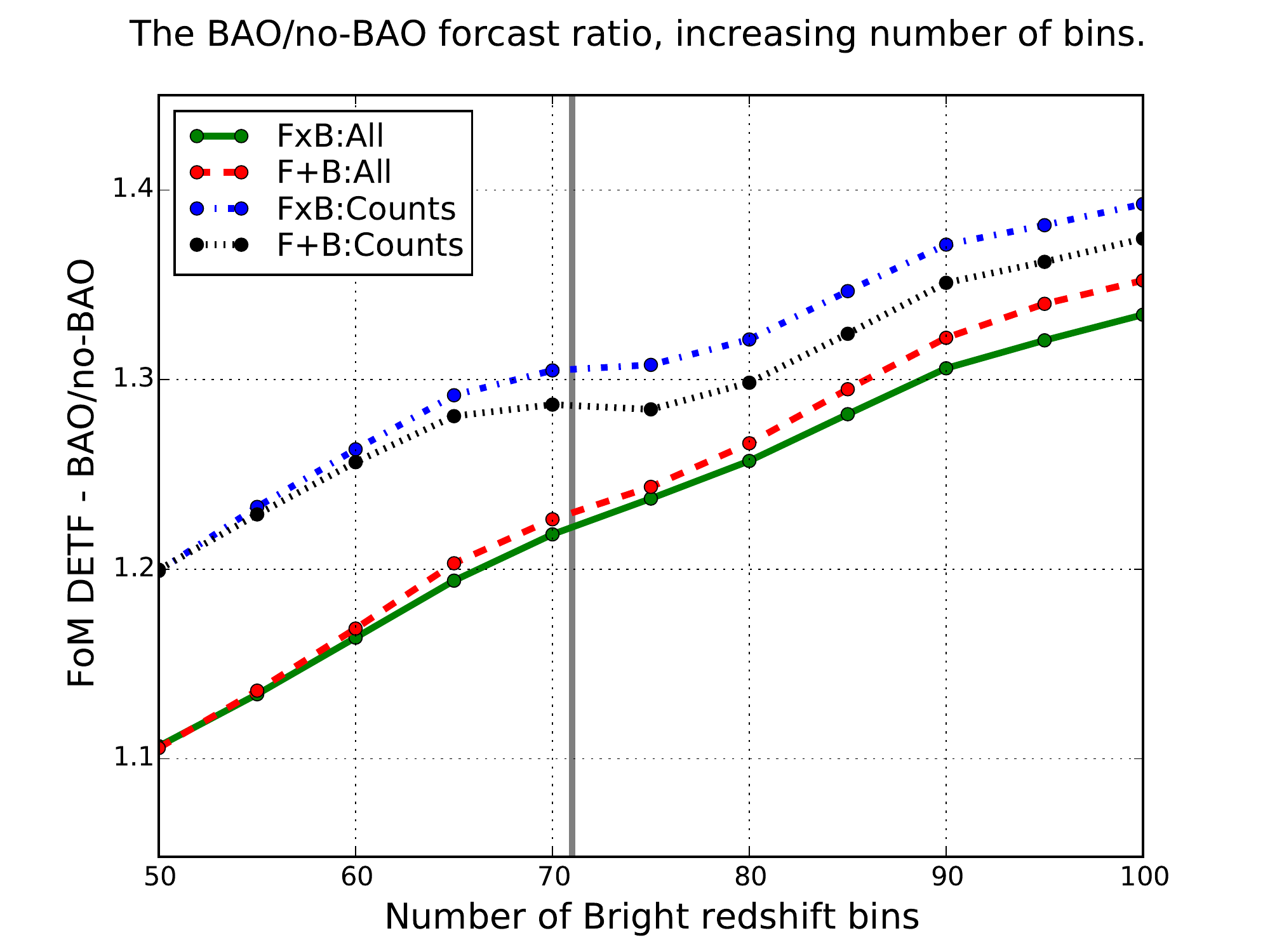}
\xfigure{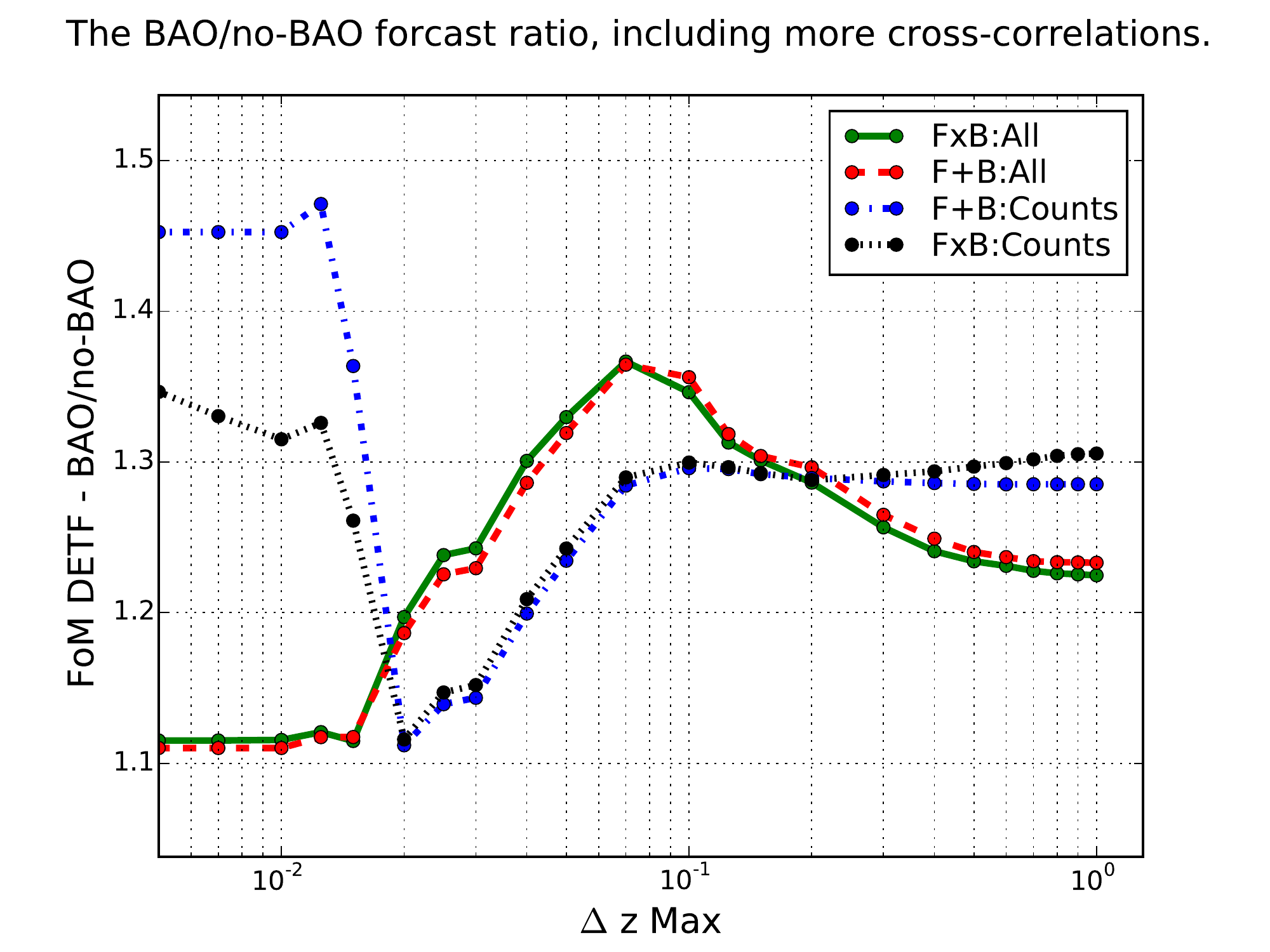}
\caption{Effect of including BAO wiggles. The ratio divides the fiducial
forecast on one removing the BAO wiggles in the Eisenstein-Hu power spectrum.
In the top and bottom panel, the $\fomdetf$ is respectively shown when varying
the number of spectroscopic redshift bins and $\dzmax$. The four lines are for
the probes FxB:All, F+B:All, FxB:Counts and F+B:Counts. A vertical line in the
upper panel corresponds to the fiducial number of spectroscopic redshift bins
(\nbright).}
\label{res_bao_comb}
\xef

Fig. \ref{res_bao_comb} show in the top panel the ratio between including or
not the BAO (BAO/no-BAO) for different number of spectroscopic redshift bins.
For the fiducial binning (\nbright\ bins), the $\fomdetf$ improve with 20-30\%, while
$\fomg$ only change with $\pm 2$\% (not shown). This trend is opposite to RSD,
discussed in subsection \ref{subsec:res_rsd}, where RSD contributed strongly to
$\gamma$ constraints ($\fomg$), but only gave minor changes to the dark energy
constraints ($\fomdetf$). Measuring $\gamma$ depends on measuring the
amplitude, while the dark energy constraints come more from the power spectrum
shape measurements (BAO position as opposed to amplitude). The RSD breaks the
degeneracy between the galaxy bias and the growth parameter ($\gamma$). On the
other hand, the BAO introduce a known distance scale, which is more suited to
measure the shape and expansion history.

The BAO peak can in configuration space be modelled by a 30 Mpc/h wide Gaussian.
For the fiducial binning (\nbright\ bins), the bin width at $z=0.5$ is 35 Mpc/h.  When
increasing the number of bins, one decrease the redshift bin width, which leads
to a more precise location of the BAO peak. Thinner bins can also, as we will
discuss, better measure the radial BAO in the cross-correlations between nearby
bins. As as result, we find the  $\fomdetf$ BAO/no-BAO ratio to around double
when using 100 instead of \nbright\ bins in the spectroscopic sample.

Fig. \ref{res_bao_comb}, bottom panel, show the BAO/no-BAO ratio when increasing
$\dzmax$. The forecast only include auto-correlations when $\dzmax = 0$, while $0.01
\leq \dzmax$ also has the cross-correlations between redshift bins. For only
auto-correlations of galaxy counts, the ratio is artificially high since the
forecast use one bias parameter per redshift bins and population. 
The "Counts" ratios therefore drops when including the cross-correlations between close 
redshift bins. Previously paper-I found a stronger BAO signal in the
cross-correlation between nearby bins than in the auto-correlation. This 
was caused by the cross-correlation selecting galaxy pairs with a given radial
distance, therefore suppressing the small-scale information. For both "All" and
"Counts" the BAO/no-BAO ratio grows from 1.1 to 1.3-1.35, showing that radial
BAO is an important contribution.

The shear-shear auto-correlations are in "All" included for all $\dzmax$
values. When including lensing the BAO/no-BAO ratio increase is higher with
$\dzmax$, since the BAO help to break degeneracies. For higher $0.1<\dzmax$
counts-shear becomes important and the ratio again decrease. Since the ratio
also decrease for a fixed bias (not shown), the decrease does not result from
count-shear providing an additional bias measurement, but from the counts-shear
signal depending more weakly on BAO. Because the galaxy counts magnification is a
weak effect (see subsection \ref{subsec:res_magn}), the lines for only counts
remains quite flat for $\dzmax>0.1$.

\subsection{Combining WL, RSD and BAO}
\label{subsec:res_gain}
\newcommand{\maincaption}[1]{
Table to compare different combinations of observables and effects. The two
tabular corresponds to #1 indicated in the upper left corner. The label column
indicates the populations in the rows (B:Bright/Spectroscopic,
F:Faint/Photometric) and using overlapping (x) or separate (-) skies. Counts
include only overdensities of number counts, while All also include galaxy
shear. The rows are divided through dashed lines in five sections. First two
sections study overlapping versus non-overlapping surveys, where the last line
is the fraction gained using overlapping surveys. Third row section present the
single populations cases (F or B). The fourth section looks at special cases,
defined in subsection \ref{subsec:assum_observ}, designed to understand which
correlations contribute most. Fifth section is the forecast for overlapping
surveys without any cross-correlations and the ratio to non-overlapping
surveys. The column "Fiducial" is the fiducial forecast, while "xBias" fixes
the galaxy bias. In the next columns are forecasts corresponding to removing
Magnification (No Magn), Weak Lensing (No WL), Redshift Space Distortions (No
RSD) and Baryonic Acoustic Oscillations (No BAO). The last three columns
include fixed bias cases.}

\begin{table*}
\begin{center}
\begin{tabular}{lrrrrrrrrrrrr}
\input{links/prod_tbl1_fomc.csv}
\input{links/prod_tbl1_fomg.csv}
\end{tabular}
\caption{\maincaption{$\fomc$ and $\fomg$}}
\label{maintable_part1}
\end{center}
\end{table*}

\begin{table*}
\begin{center}
\begin{tabular}{lrrrrrrrrrrrr}
\input{links/prod_tbl1_fom.csv}
\input{links/prod_tbl1_fomdetf.csv}
\end{tabular}
\caption{Same as Table \ref{maintable_part1} for
$\fom$ and $\fomdetf$}
\label{maintable_part2}
\end{center}
\end{table*}

Previous subsections studied in detail the separate benefits of the
covariance, RSD, WL and BAO. This subsection build on those and compares the
relative impact of each physical effect, including knowledge of galaxy bias and
sample variance cancellations. The main results are presented in four tabulars,
corresponding to $\fomc$, $\fomg$, $\fomdetf$ and $\fom$. For layout reasons,
$\fomc$ and $\fomg$ are shown in Table \ref{maintable_part1}, while $\fomdetf$
and $\fom$ are included in Table \ref{maintable_part2}. Each row corresponds to
a different probe and dashed lines divide the rows into five sections. The
first two sections quantify the benefit of overlapping photometric and
spectroscopic galaxy surveys. In the third section we study the single
population, while the fourth and fifth section present special cases (see
Table \ref{notation_one}). On the columns are the forecast for a free/fixed
galaxy bias and when removing different effects.

The first three rows of Table \ref{maintable_part1} show the forecast for
FxB:All, F+B:All and the same-sky benefit: FxB/F+B. For $\fomc$ we find a 50\%
same-sky gain, which corresponds to 30\% increase in area (Eq. \ref{areagain})
The origin of this benefit in $\fomc$ is explained in a companion paper
\citeind{sameskyX}. Here we focus on describing the different results in
detail. In the dark energy FoMs ($\fom$ and $\fomdetf$) the benefit is similar,
while the $\fomg$ ratio is 1.1, which corresponds to a 20\% larger area. While
the details differ, for galaxy counts and shear we find similar benefits from
overlapping photometric and spectroscopic surveys. In general, the absolute
numbers in the forecast presented depends strongly on the parameterization of
the galaxy bias. For example, exact knowledge of bias would increase FxB:All
and F+B:All with 6.0x and 7.9x respectively. Details on the galaxy bias is
studied in paper-III.

For the four defined FoMs, the survey overlap is more important when
marginalising over the bias. One example is the fiducial column, where the
FxB:All/F+B:All ratio for $\fomc$ decrease from 1.5 to 1.1 when fixing the
galaxy bias. Also, we see a lower gain from overlapping surveys when
including RSD or BAO. Those effects break degeneracies between the galaxy bias
and cosmology, which increase the single population cases and therefore reduce
the importance of overlapping surveys. Without lensing and for a fixed bias,
we find overlapping surveys contribute negatively. This is because there are no
additional counts-shear cross-correlations and the reduced sampling variance
only works with a free bias \citeind{sameskyX}.

If we focus on the FxB:All case, we see that the galaxy bias is the effect
causing larger impact in the FoMs: fixing bias increases $\fomc$ by a factor
x6.0. For free bias, the most important probe is WL (x4.8), then RSD (x2.1) and
finally BAO (x1.3). If we look at $\fomg$ the order is preserved WL (x1.9) and
RSD (x1.7), while BAO has no impact on $\fomg$. For $\fom$ or $\fomdetf$ we see
that WL is still the most important effect, but here BAO is more relevant that
RSD, which makes sense as the former measures distances, while the later
measures growth, which is more relevant for $\fomg$. When bias is known (xBias)
the $\fomc$ in overlapping surveys (FxB) the relative impact of other effects
is smaller, but we have similar hierarchy of tendencies: WL (x3.6), RSD (x1.3)
and BAO (x1.04). For non-overlapping surveys (F+B) and free bias the gain is
smaller and both RSD (x2.2) and BAO (x1.4) become more important relative to WL
(x4.0).

Second section of rows is the forecast and ”overlapping skies ratio” only using
galaxy counts. The constraints without galaxy shear is lower, with
FxB-All/F+B-All being 1.4, 1.2, 1.2 and 1.2 for $\fomc$, $\fomg$, $\fom$ and
$\fomdetf$. Fixing the bias of the fiducial case (xBias), the table shows how
the ”Improvement” ratio becomes close or slightly below unity, meaning all the
benefit of cross-correlating the galaxy counts comes from measuring the galaxy
bias. This is different from "All" where $\fomc$ and $\fom$, which depends both
on the DE parameters and $\gamma$, improve also for a fixed bias. With shear
the overlapping surveys include an additional counts-shear signal, while for
counts the only benefit comes from better bias measurements. Not including RSD
degrade the galaxy bias and growth determination, which is compensated when
overlapping surveys which include additional cross-correlations and sample variance
cancellations to better measure the galaxy bias.

The third section shows the single population constraints, with F:All and
B:All respectively being the optimal (both counts and shear) constraints for
the Faint and Bright population. Below are the cases F:Counts and B:Counts,
which only include galaxy counts. Weak lensing is the main contribution to the
Faint sample, while the Bright sample constraints are driven by galaxy
clustering and RSD. This can be seen by comparing "All" and "Counts". The
ratios F:(All/Counts) and B:(All/Counts) are respectively 38 and 1.4 for
$\fomc$. In our forecast the photometric sample use photo-z redshifts, which
drastically reduce the contribution from RSD and intrinsic galaxy counts
cross-correlations between redshift bins. The low F:Counts constraints can
suggest using less the current 12 Faint bias parameters (\nbright\ for the Bright),
but the constraints would still be relatively low. Also a spectroscopic survey
can not measure shear (in B:All), but the Bright includes shear when being a
subset of an overlapping photometric survey.

The fourth section show the forecast for FxB:All when not including various
counts-shear cross-correlations (see notation Table \ref{notation_one}). The
counts-shear signal is important for the combined constraints. Comparing
FxB:All to \gkthree:All, we see how $\fomc$ almost double (2.2x free bias) when
including the counts-shear correlations, while the $\fom$ and $\fomdetf$
greatly improve (86\% and 69\%, free bias). For $\fomg$ the change is smaller
and the increase is respectively 22\% and 1\% for a free and fixed bias.
Including only either counts-shear cross-correlations of spectroscopic
(\gkone:All) or photometric (\gktwo:All) galaxy counts give comparable
constraints. Removing the counts-shear cross-correlations altogether
(\gkthree:All) leads to a drastic drop. We therefore conclude the
counts-shear cross-correlations are important, but multiple populations include
redundant information.

Last row section study the direct same-sky improvement from overlapping volumes.
In addition to the extra-correlation, the overlapping volumes increase the
covariance between different galaxy samples probing the same dark matter
fluctuations. The additional covariance result in a larger covariance between
the bias parameters. When marginalising over the bias, this can improve the
constraints (\cite{sameskyX}). The volume effect is quite small for only galaxy
counts, since the constraints are mainly from the Bright/spectroscopic sample.
When lensing is included the overlapping volumes measure the Faint bias and
therefore improve constraints through the counts-shear cross-correlations.

\subsection{Magnification}
\label{subsec:res_magn} In Table
\ref{maintable_part1} and \ref{maintable_part2}, the "No Magn" column remove
the effect of magnification. Weak gravitational lensing increase galaxy fluxes,
altering the galaxies entering into a magnitude limited sample.  Foreground
matter also magnify the area, which change the observed galaxy densities. These
two effects together is the weak lensing magnification with number counts.
Removing magnification is done by setting the magnification slope to zero (see
Fig. \ref{mag_slopes}). For FxB:All the $\fomc$, $\fom$, $\fomdetf$ improve
1\% from magnification, while the $\fomg$ is close to zero ($< 0.1\%$). The
improvement is significant when only including galaxy counts. For FxB:Counts
the magnification contributes to $\fomc$, $\fomg$, $\fom$ and
$\fomdetf$ with respectively 13,10,5 and 4\%. 

In the previous paper \citedir{gazta}, we studied the impact of magnification
and recently \citedir{duncanmag} confirmed those findings. One source of
confusion was the notation "MAGN", which denoted magnification combined with
galaxy clustering. While stating the galaxy clustering was the main source for
the constraints, the misleading labels and partly unclear text lead some
readers to believe magnification had a more central role. In this paper the
forecasts of galaxy counts is simply labeled "Counts" and includes galaxy
clustering, RSD and magnification. Unlike the previous article, this paper also
discuss the effect of magnification also when including shear. We note that
magnification is potentially more effective if marginalising of additional
systematics (like photo-z outliers). A detailed study of constraining lensing
systematics when combining with magnification is left for future work.

%% file: concl.tex
The effects of galaxy clustering, redshift space distortions (RSD), weak
lensing (WL) and BAO are presented for 2D angular cross-correlations of galaxy
counts and shear. Building on paper-I, which presented the modelling, this
paper use the Fisher matrix formalism to estimate dark energy and growth rate
($\gamma$) constraints for photometric and spectroscopic surveys. The forecast
use two galaxy populations, one photometric (F) and one spectroscopic (B), and
analyze the spectroscopic survey in 72 narrow redshift bins to capture the radial
information. All possible cross-correlations between galaxy counts and shear
are included. In this paper we focus on the relative benefit of different
correlations and effects as non-linear contributions, the Limber approximation,
the covariance, RSD, BAO and magnification. Details on the forecast assumptions
and nomenclature can be found in section \ref{assumpt}.

To prevent entering into the strong non-linear regime, subsection
\ref{subsec_nonlin} defined a criteria for which correlations to include. In
appendix \ref{app_nonlin} we study the non-linear effect, finding our $k_{max}$
cut to be reasonable. The next subsection(\ref{subsec:res_radial}) we compare
the benefit of different correlations. The effect of galaxy clustering, RSD and
WL all enters in the 2D correlations. To investigate their relative impact we
introduce the variable $\dzmax$, which limits the maximum distance between the
mean of the two redshift bins in a correlation. For $\dzmax=0$ only the
auto-correlations are included, then $\dzmax \approx 0.02$ also include
cross-correlations between nearby bins in the spectroscopic sampling, while
counts-shear and magnification only enters for higher $\dzmax$. From plotting
the figures of merit (FoM) as a function of $\dzmax$, we show how the different
correlations contribute. This includes the cross-correlations between nearby
bins in the spectroscopic sample.

The Limber approximation is widely used to simplify the calculation of the
galaxy clustering in 2D correlation. As shown in paper-I, the Limber
approximation only works in thick redshift bins. For narrow bins, which we need for
the spectroscopic sample, the Limber approximation breaks down and incorrectly
estimate zero cross-correlation for close redshift bins. Subsection
\ref{subsec:res_limber} show the effect on the forecast. The Limber
approximation overestimate the amplitude, therefore reducing the impact of
shot-noise. The exact calculations give larger errors in $\gamma$ than the
Limber approximation. More importantly, the cross-correlations of galaxy counts
in nearby bins are effective in constraining dark energy. Since these are zero
for the Limber approximation, it leads to the exact calculation giving stronger
dark energy constraints.

The subsection \ref{subsec:res_resolution} study the effect of including more
bins in the spectroscopic sample. For as increasing number of bins, we find the
forecast is saturating from a higher covariance. This result has previously
been expected, but not shown due to technical issues with a large number of
bins \citeind{asorey,classgal}. In the spectroscopic sample, for most bin 
configurations (less than 160 bins) the covariance improve the forecast. This
effect come from the covariance between redshift slices, which reduce the
sample variance similar to a multi-tracer analysis (see \citedir{sameskyX}).
Lastly, we find our same-sky result to be stable over a larger number of
spectroscopic bins.

In subsection \ref{subsec:res_rsd} we show how RSD break the degeneracy
between the galaxy bias and f, which result in better $\gamma$ constraints.
Similar to the Limber subsection, the RSD/no-RSD ratio is shown as a function
of $\dzmax$. The RSD effect suppress the galaxy counts cross-correlations of
close redshift bins, which reduce their constraint. For more spectroscopic
redshift bins (100 instead of 72), the RSD impact increase for the combined
photometric and spectroscopic surveys. This subsection also discuss how RSD
minimally impacts the signal, but decrease the counts-shear constraints through
increasing galaxy counts auto-correlation which enters in the error estimate.

Opposite to RSD, the BAO contribute significantly to dark energy constrains,
but only has a minor impact on $\gamma$ constraints. Subsection
\ref{subsec:res_bao} show, similar to previous subsection, the BAO/non-BAO
ratio when varying $\dzmax$ and the number of spectroscopic redshift bins. The
no-BAO forecast is estimated by using the Eisenstein-Hu power spectrum without
the BAO wiggles. In paper-I we showed that the cross-correlations between
narrow and close redshift bins has a higher (radial) BAO contribution compared
with the auto-correlations. This is reflected in dark energy constraints
depending stronger on BAO when these (radial) cross-correlations are included.

Subsection \ref{subsec:res_gain} include the four main forecast tabulars (Table
\ref{maintable_part1},\ref{maintable_part2}), each corresponding to a different
FoM. The different rows corresponds to probes, defined by which galaxy
populations (photometric, spectroscopic) included, if the surveys overlap
observable (counts, shear) are used and if some cross-correlations are removed.
Columns correspond to modifying some effect, as removing WL, magnification, RSD
or BAO and if marginalising or fixing the galaxy bias. For the combined
overlapping photometric and spectroscopic survey, the bias is the physical
effect with largest impact on the $\gamma, w_0, w_a$ combined figure of merit
$\fomc$. When marginalising over the bias (free bias), the next effective in
relative importance is WL (factor of 4.8), RSD (factor 2.1), BAO (factor 1.3)
and magnification (1$\%$). Magnification is discussed separately in subsection
\ref{subsec:res_magn}, comparing with the literature and clarifying the
difference in notation with \citedir{gazta}.

Two photometric (F) and spectroscopic (B) surveys increase $\fomc$ equivalent to
30\% larger area when overlapping. The benefit is smaller for a known galaxy
bias. Overlapping surveys (FxB) improve the constraints for two reasons.
Additional cross-correlations for overlapping surveys can explain part of the
gain. One can cross-correlate the F and B galaxy counts, and also foreground
spectroscopic counts with the shear from the photometric survey. The second
contribution is the additional covariance since the overlapping surveys (F and
B) trace the same matter fluctuation. This advantage of two galaxy population
has already been show in 3D P(k) \citeind{mcdonseljak} and 2D correlations
\citeind{asorey2}. Here we extend those findings to WL and RSD for the
combination of the F and B samples. \citedir{sameskyX} explain these effects in
more detailed and compares our forecast to other analysis in the literature.

%% file: acknow.tex
We would like to thank the group within the DESI community looking at
overlapping surveys. M.E. wish to thank Ofer Lahav, Henk Hoekstra and Martin
Crocce in his thesis examination panel, where these results were discussed. E.G
would like to thank Pat McDonald for exploring and comparing results. Funding
for this project was partially provided by the Spanish Ministerio de Ciencia e
Innovacion (MICINN), project AYA2009-13936 and AYA2012-39559,
Consolider-Ingenio CSD2007- 00060, European Commission Marie Curie Initial
Training Network CosmoComp (PITN-GA-2009-238356) and research project 2009-
SGR-1398 from Generalitat de Catalunya. M.E. was supported by a FI grant from
Generalitat de Catalunya. M.E. also acknowledge support from the European
Research Council under FP7 grant number 279396.

%% file: nonlin.tex
\xbf
\xfigure{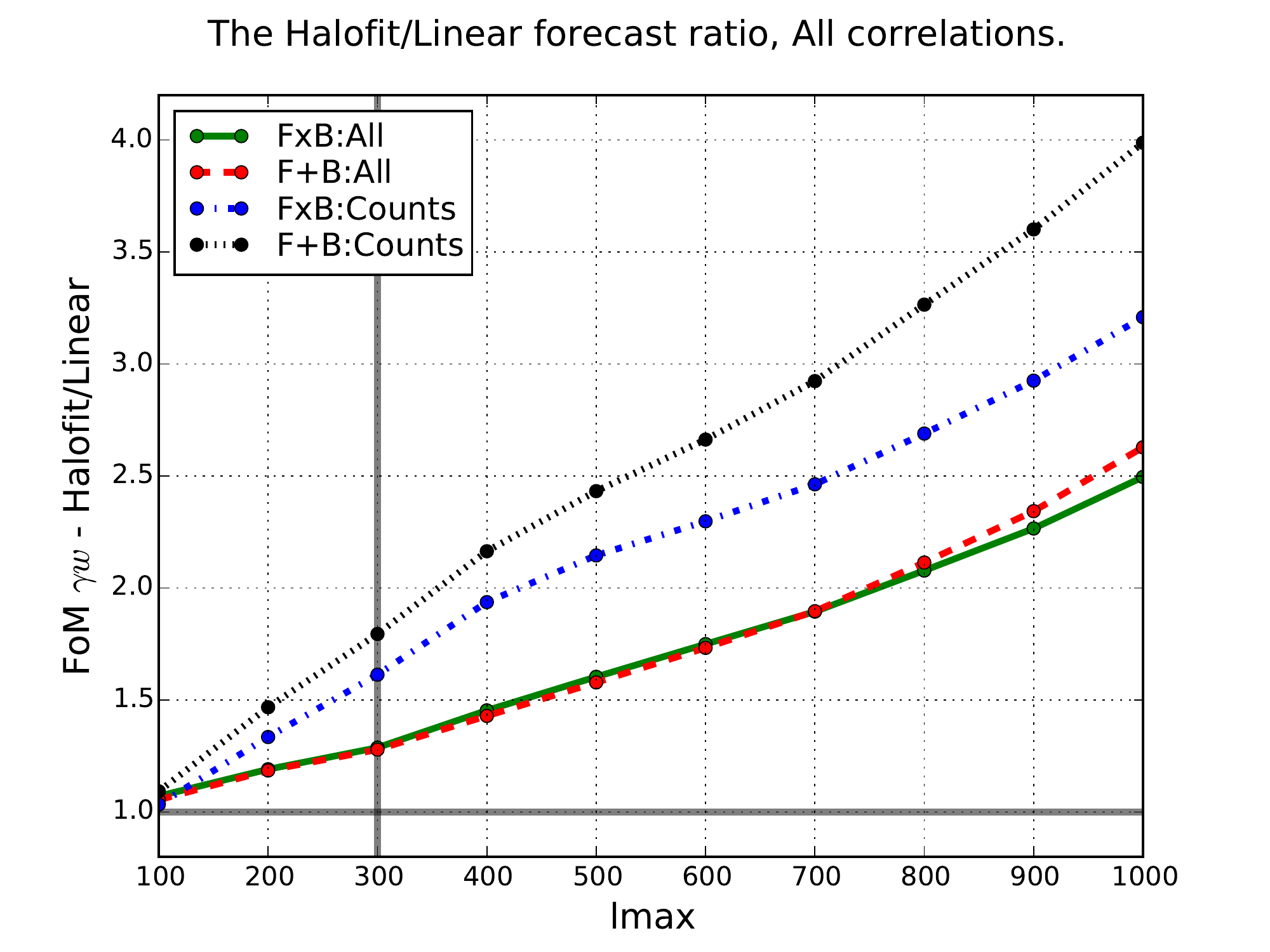}
\xfigure{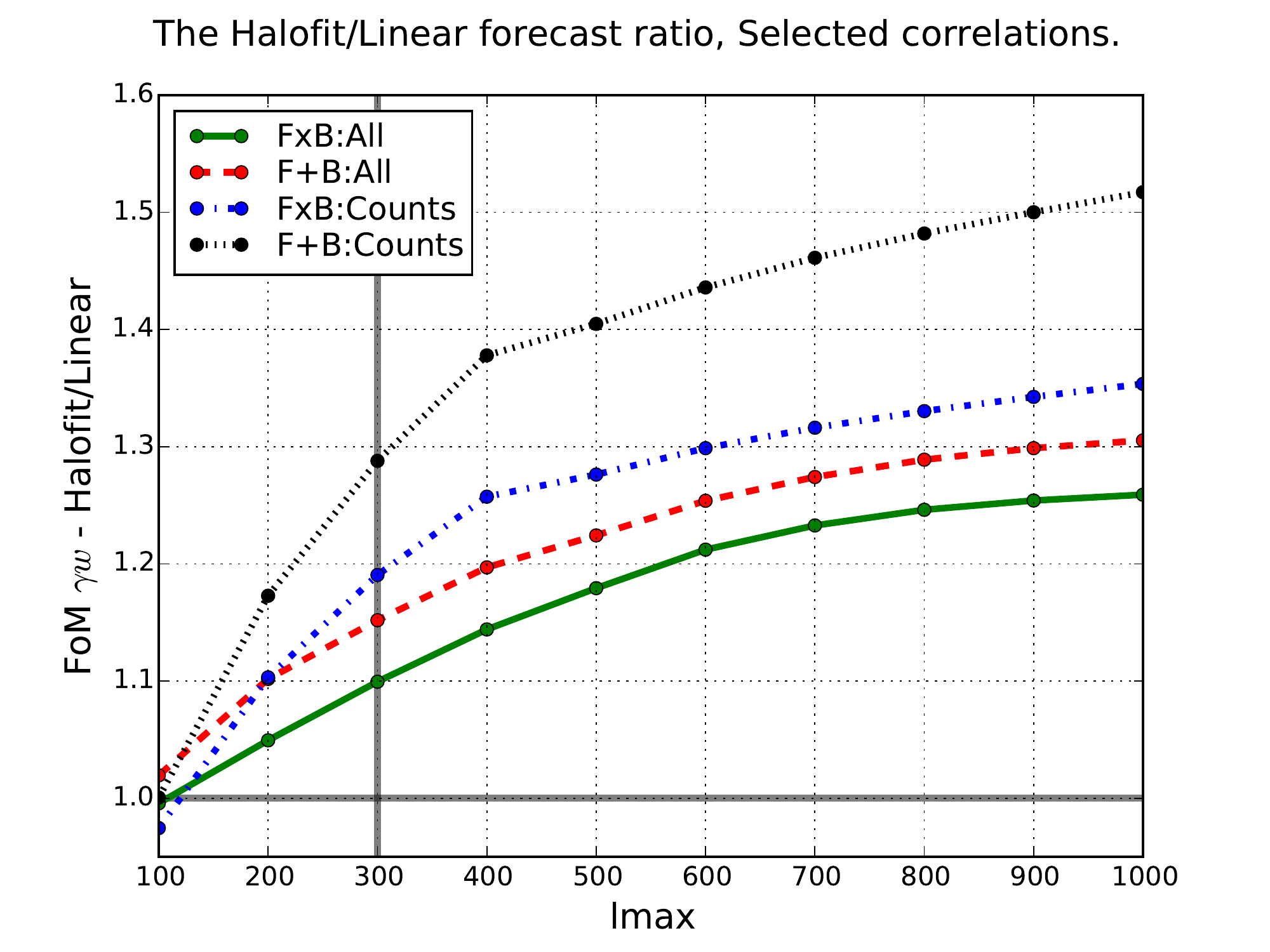}
\caption{Forecast ratio between including the Halofit contributions or only the
linear (Eisenstein-Hu 2003) power spectrum. The figure of merit is $\fomc$ and
the lines corresponds to FxB:All, F+B:All, FxB:Counts and F+B:Counts. In the
top/bottom panel, the ratios include/exclude correlations entering into
non-linear scales ($k>k_{max}$). A vertical line at $l_{max} = 300$ mark the
fiducial value.}
\label{halofit_ratio}
\xef

The fiducial forecast include Halofit-II, use $10 \leq l \leq 300$ and remove
correlations entering into non-linear scales (subsection \ref{subsec_nonlin}).
To test the impact of non-linear scales, Fig.\ref{halofit_ratio} show the 
$\fomc$.  ratio between including non-linear P(k) (Halofit II) and only the
corresponding linear spectrum (from Eisenstein \& Hu 2003, EH). On the x-axis
is the maximum multipole included, $l_{max}$, which fiducially is $l_{max} =
300$. The top panel shows this ratio using all correlations until the $l_{max}$
cut. As expected, the ratio increase with $l_{max}$ and it is largest when
only including galaxy counts.

The bottom panel illustrate the effect of an additional cut to remove
correlations entering into non-linear scales $k>k_{max}$. For $l_{max}=300$ the
$\fomc$ including Halofit is for FxB:Counts respectively 30\% and 80\% higher
than EH only when removing or using all correlations. With increasing
$l_{max}$, the ratio with "All" correlations grows quite linearly, while for
the "selected correlations" the ratios flattens. The $\fomdetf$ and
$\fomg$ (not shown) follow the same pattern, but with smaller ratios. For
$\fom$ (not shown) the ratios are even flatter at high  $l_{max}$  and none of
the probes cross the line 1.2. The fiducial forecast include Halofit, but we
limit the correlations included to not become too sensitive to assumptionts on
the non-linear scales. In addition to the non-linear matter power spectrum,
these scales could require a scale dependent galaxy bias. Note that these 
results are on the forecasted accuracy, which mainly depend on the observable
derivative with respect to cosmology. The cut in non-linear scales for not 
biasing a parameter fit (precision) can be different and is not considered 
here.

%% file: contours.tex
\xbf
\xfigure{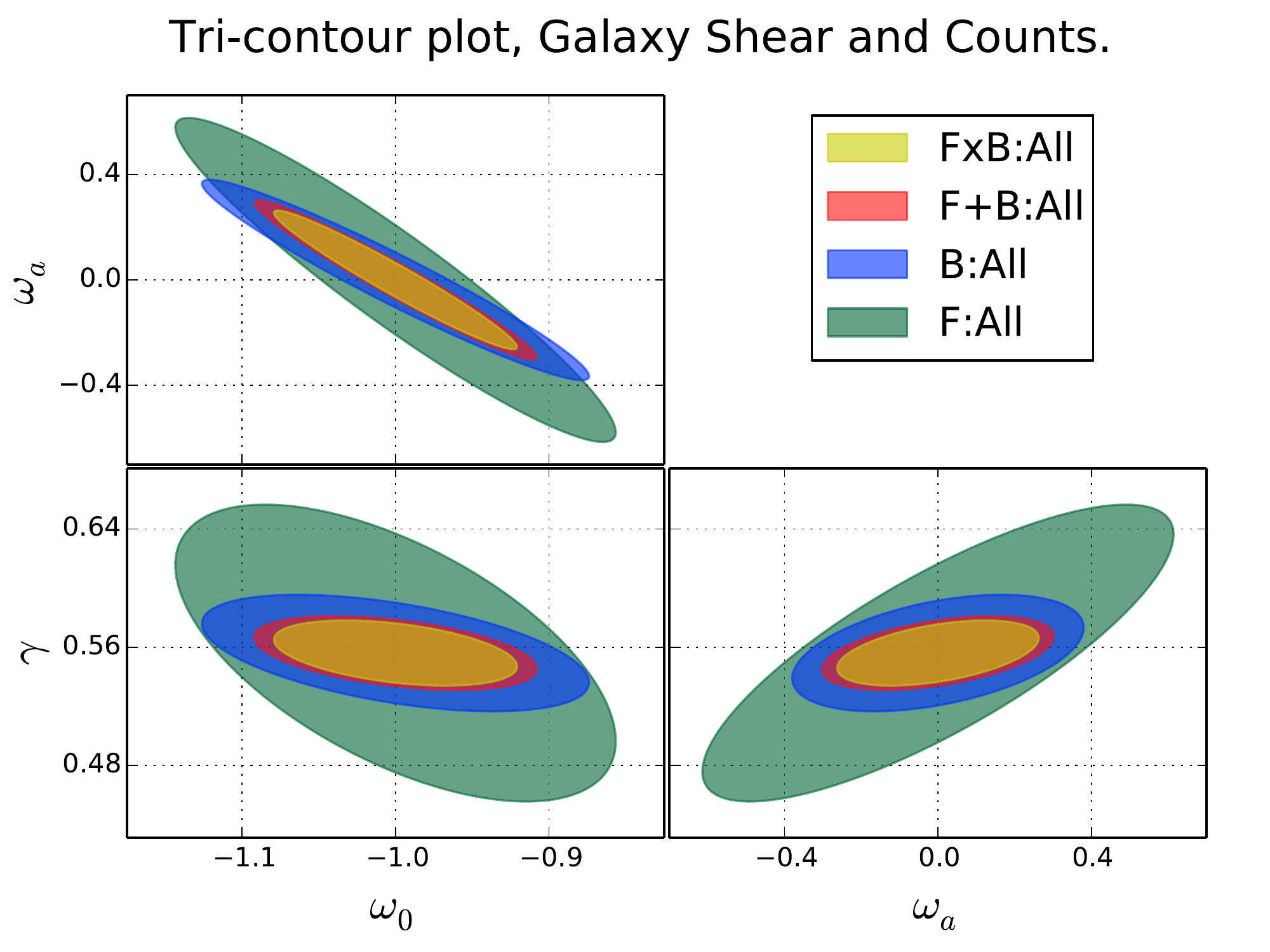} 
\xfigure{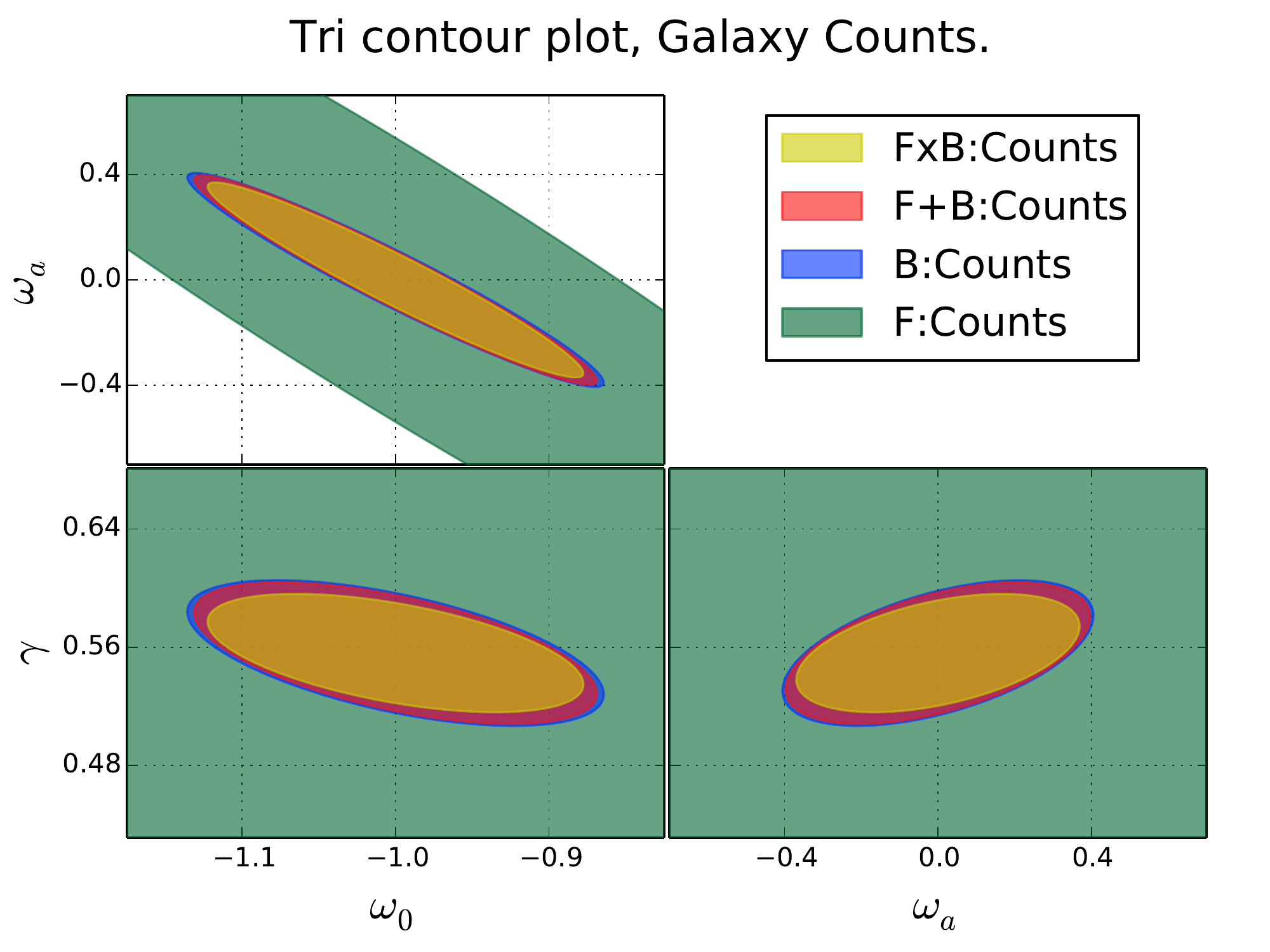} 
\caption{Contour plots of $w_0$, $w_a$ and $\gamma$. The three sub-plots show
the Fisher matrix 1-$\sigma$ contours, marginalizing over the DETF parameters
and galaxy bias. The top panel show contours for FxB:All, F+B:All, B:All and
F:All, while the bottom panel only include galaxy counts.}
\label{res_contour_main}
\xef

Fig. \ref{res_contour_main} shows the 1-$\sigma$ contours for $w_0$, $w_a$ and
$\gamma$. The top panel show for "All" and the combination FxB, F+B, F and B.
One can see some trends also present in the tables. The combination F+B:All,
combining shear and galaxy counts from separate surveys, is more powerful than
analyzing the survey separately. The factor of 1.5 improvement of FxB:All over
F+B:All corresponds to the difference between the two inner ellipses. On the
bottom is a similar plot for the galaxy counts. Using equal scales allow us to
directly compare the constraints, but at the expense of the F:Counts contours
being plotted beyond the borders. For the galaxy counts the Bright population
completely dominates, even if the Bright sample includes more bias parameters. 

\xbf
\xfigure{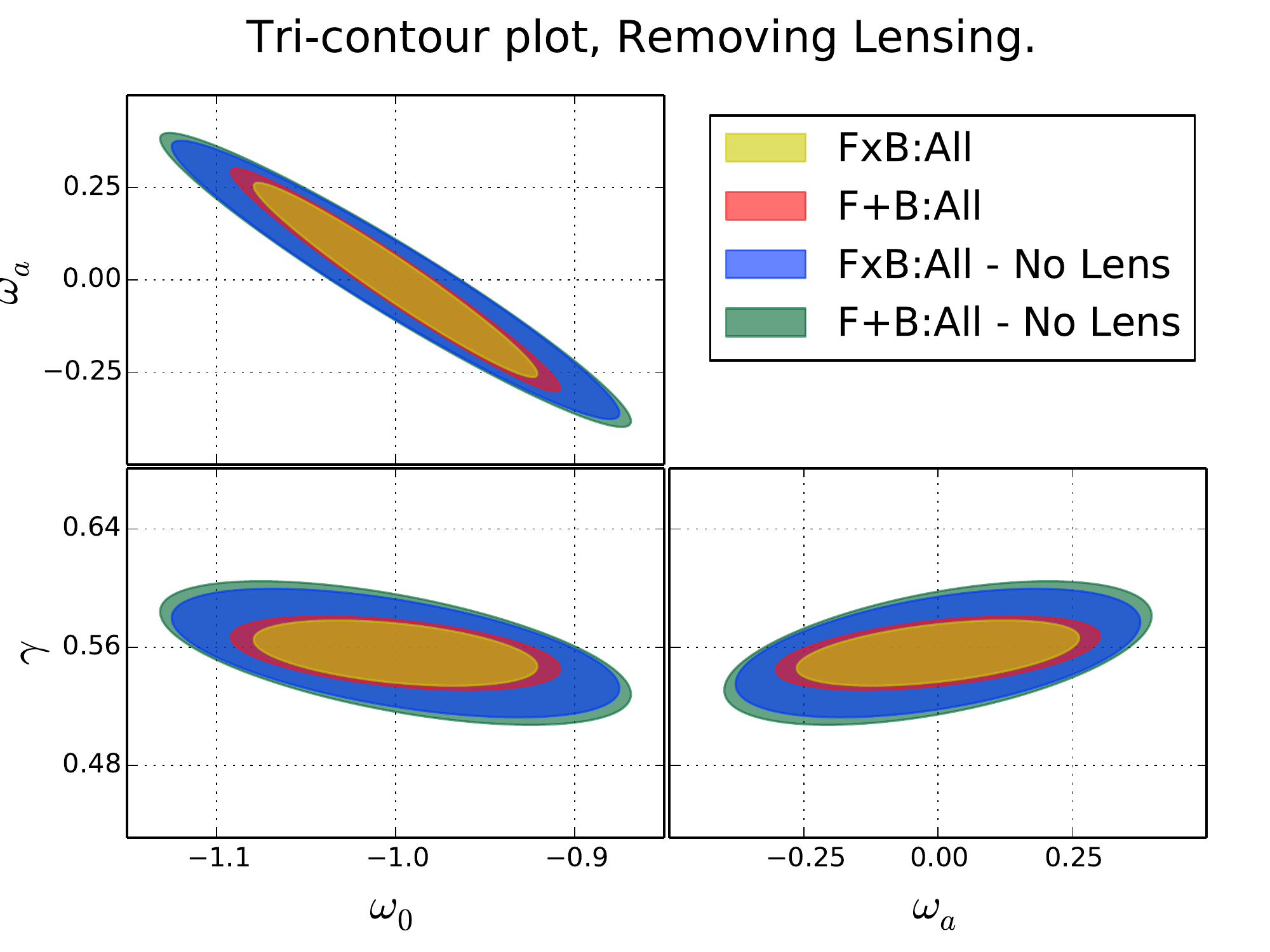} 
\xfigure{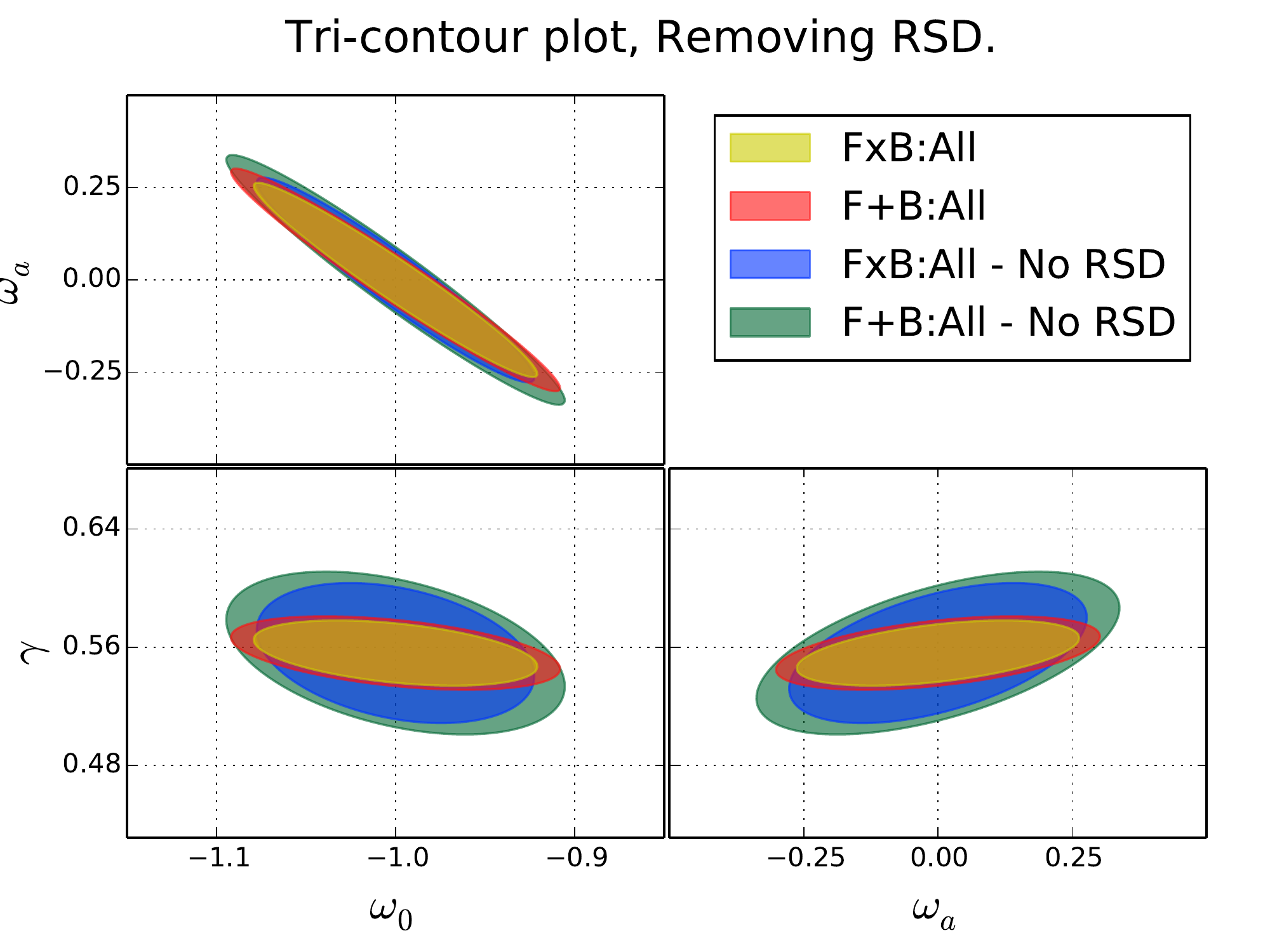} 
\caption{Contour plots of $w_0$, $w_a$ and $\gamma$. The three sub-plots show
the Fisher matrix 1-$\sigma$ contours, marginalizing over the DETF parameters
and galaxy bias. In the upper plot, two ellipses are the fiducial FxB:All and
F+B:All, while two remove Weak Lensing observables. The bottom plot similarly
show the fiducial FxB:All and F+B:All, and then two contours in real space.}
\label{res_contour_lensrsd}
\xef

Last Fig. \ref{res_contour_lensrsd} looks at the effect of
removing WL and RSD. The equivalent Magnification and BAO plots are not
included since those effects are weaker, which results in less difference
between the ellipses. The top panel shows FxB:All and F+B:All in a tri-contour
plot, with and without WL (fiducial and "No Lens"). The Weak Lensing improve
the constraints on all three parameters included in the contour plots.
Comparing the FoMs in Table \ref{maintable_part1} and \ref{maintable_part2},
one see same-sky benefit of FxB:All is actually higher when including lensing.

In the lower panel is similar plot, instead with two contours calculated with
and without RSD (fiducial and "No RSD"). While the RSD impact the parameter
constraints different, the margins are exactly equal so one can visually
compare the effects. The RSD is contributing strongly to measuring $\gamma$ and
less to $w_0$ and $w_a$. One also see the same trend in the Tables
\ref{maintable_part1} and \ref{maintable_part2}. There the RSD improve $\fomc$,
$\fomg$ and $\fom$ which depends on $\gamma$, while not $\fomdetf$ where
$\gamma$ is fixed. The difference between the contours in bottom panel show if
RSD increases or decreases the importance of overlapping surveys. Including
RSD, looking at the numerical values in the table, slightly reduce the benefit
of overlapping galaxy surveys.